%% file: acl_latex.tex
\pdfoutput=1

\documentclass[11pt]{article}
\usepackage{acl}
\usepackage{times}
\usepackage{latexsym}
\usepackage[T1]{fontenc}
\usepackage[utf8]{inputenc}
\usepackage{microtype}
\usepackage{inconsolata}
\usepackage{graphicx}
\usepackage{comment}
\usepackage{amsmath}
\usepackage{comment}
\usepackage{fontawesome5}
\usepackage{boldline}
\usepackage[ruled,vlined,noend]{algorithm2e}
\usepackage{bm}
\usepackage{lipsum}
\usepackage{multirow}
\usepackage{amsfonts}
\usepackage[most]{tcolorbox}
\usepackage{listings}
\usepackage{color}
\usepackage{colortbl}

\usepackage{booktabs}  

\definecolor{dkgreen}{rgb}{0,0.6,0}
\definecolor{mauve}{rgb}{0.58,0,0.82}
\definecolor{listinggray}{rgb}{0.5, 0.5, 0.5}

\title{CodeReviewQA: The Code Review Comprehension Assessment for Large Language Models}

\author{
  \textbf{Hong Yi Lin$^\spadesuit$, Chunhua Liu$^\spadesuit$, Haoyu Gao$^\spadesuit$} \\ \textbf{Patanamon Thongtanunam$^\spadesuit$, Christoph Treude$^\clubsuit$}\\
  $^\spadesuit$The University of Melbourne, $^\clubsuit$Singapore Management University \\
  \texttt{\{tom.lin1, chunhua.liu1, haoyu.gao3, patanamon.t\}@unimelb.edu.au} \\
  \texttt{ctreude@smu.edu.sg}
}

\begin{document}
\maketitle

\begin{abstract}
State-of-the-art large language models (LLMs) have demonstrated impressive code generation capabilities but struggle with real-world software engineering tasks, such as revising source code to address code reviews, hindering their practical use. 
Code review comments are often implicit, ambiguous, and colloquial, requiring models to grasp both code and human intent. 
This challenge calls for evaluating large language models' ability to bridge both technical and conversational contexts. 
While existing work has employed the automated code refinement (ACR) task to resolve these comments, current evaluation methods fall short, relying on text matching metrics that provide limited insight into model failures and remain susceptible to training data contamination.
To address these limitations, we introduce a novel evaluation benchmark, \textbf{CodeReviewQA} that enables us to conduct fine-grained assessment of model capabilities and mitigate data contamination risks.
In CodeReviewQA, we decompose the generation task of code refinement into \textbf{three essential reasoning steps}: \textit{change type recognition} (CTR), \textit{change localisation} (CL), and \textit{solution identification} (SI). Each step is reformulated as multiple-choice questions with varied difficulty levels, enabling precise assessment of model capabilities, while mitigating data contamination risks. 
Our comprehensive evaluation spans 72 recently released large language models on \textbf{900 manually curated, high-quality examples} across nine programming languages. 
Our results show that CodeReviewQA is able to expose specific model weaknesses in code review comprehension, disentangled from their generative automated code refinement results.~\footnote{Data Availability:~\href{https://huggingface.co/datasets/Tomo-Melb/CodeReviewQA}{https://huggingface.co/datasets/Tomo-Melb/CodeReviewQA}}
\end{abstract}

\begin{figure}[ht]
    \centering
    \includegraphics[width=\columnwidth]{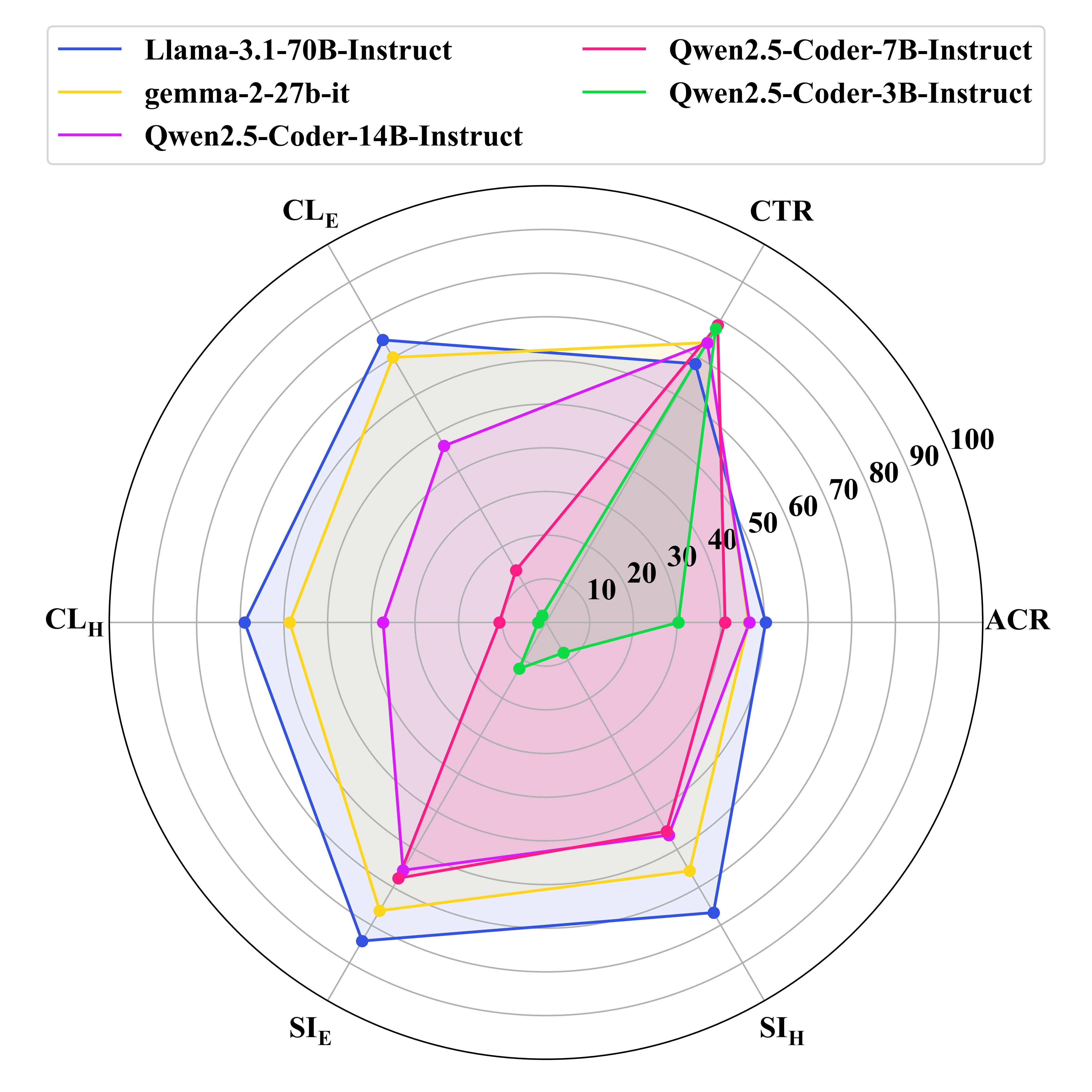}
    \caption{\textbf{CodeReviewQA} results (\%) of the top performing model (per scale class) based on ACR Accuracy. 
    \textbf{ACR:} Automated Code Refinement, \textbf{CTR:} Change Type Recognition, \textbf{CL:} Change Localisation, \textbf{SI:} Solution Identification, \textbf{E:} Easy, \textbf{H:} Hard.
    }
    \label{fig:pareto}
\end{figure}

\section{Introduction}

The proficiency of state-of-the-art large language models (LLMs) in code generation has garnered significant attention~\cite{bigcodebench}, demonstrating their ability to follow explicit instructions to author code. However, their competency in real-world software engineering environments remains limited \cite{Pornprasit2024GPT3.5}, particularly in collaborative tasks involving colloquial and complex forms of communication.  A quintessential example is code reviewing, where review comments~\cite{evacrc, language_matters} represent natural communication between developers with a shared mental model, often resulting in under-specified, ambiguous, and implicit expressions of intent. For example, this comment \textit{``For all of the fuzz tests, does it make sense to have versions for `len\_prefixed' both `true' and `false'?''} is a rhetorical question that expresses an intended code change without explicitly detailing the literal change requirement and instructions. 

As a result, the ability to resolve code review comments not only requires proficiency in understanding and generating code, but also the ability to comprehend the communicative intent behind natural language code reviews in relation to the source code that they address. Therefore, assessing how LLMs resolve code review comments serves as a crucial testbed for their proficiency in understanding and following implicit, conversational instructions in software development. Success in this domain would significantly advance automated software development assistance, potentially reducing developer workload and improving code quality.

To evaluate model capability in resolving code review comments, prior work has explored the \textit{automated code refinement} (ACR) task using both small scale neural language models \cite{tufano2022, autotransform} and LLMs~\cite{guo2024exploring,Pornprasit2024GPT3.5}, aiming to automatically revise source code based on code review comments. 
While these efforts have advanced this direction, several critical challenges remain unaddressed. 
Firstly, current automatic evaluation approaches rely heavily on metrics such as exact match and BLEU~\cite{bleu}, which merely capture surface-level token similarities between the generated output and the ground truth, without reflecting intermediate comprehension capabilities. 
Secondly, as these evaluation benchmarks typically use popular GitHub projects, they risk data contamination from training data in LLMs~\cite{llmthreats}, potentially obfuscating true model capabilities. 
As a result, there are currently no suitable approaches for assessing the capabilities of state-of-the-art LLMs in automated code refinement.

To address these challenges, we introduce a novel evaluation benchmark that enables comprehensive assessment of automated code refinement capabilities. Our benchmark decomposes the original one-step generative task into three underlying reasoning steps: \textit{change type recognition} (CTR), \textit{change localisation} (CL), and \textit{solution identification} (SI). 
These components represent essential cognitive processes required for understanding code review comments, and subsequently generating the required code revision.
By reflecting explicit intermediate reasoning steps, our benchmark provides fine-grained feedback of model failures to support future model development.

To mitigate potential data contamination, we formulate each reasoning step as a multiple-choice question answering (MCQA) probe with synthetic answers. 
This approach transforms the original task into unfamiliar formats with unseen solutions, demanding proficiency in code review comprehension, rather than sequence memorisation of contaminated data~\cite{cleaneval}. 
Furthermore, we leverage MCQA's flexibility to introduce distractor variation strategies, enabling assessment of model understanding across different difficulty levels.

To avoid the pervasive issue of noisy examples present in past benchmarks~\cite{tufano2024code} and ensure high-quality evaluation data, we manually verify and curate 900 valid code refinement examples that cannot be automated by traditional software engineering tools. These examples are sourced from 199 repositories, reflecting nine of the most popular programming languages on GitHub. Finally, we evaluate 72 state-of-the-art code-intelligent LLMs, providing an extensive benchmark to facilitate future research.
Our contributions can be summarised as follows:

\textbf{1) CodeReviewQA Benchmark.} 
The first ACR evaluation approach to include intermediate reasoning probes for detailed feedback on code review comprehension capabilities of LLMs.
It is also the first approach to counteract the effects of data contamination in ACR assessment, allowing for the reuse of existing code review data in LLM evaluation.
The benchmark consists of four tasks (three MCQA probes and one generative task) in total.

\textbf{2) Clean Code Review Evaluation Set.} 
The first code review evaluation set that has been completely manually verified for examples that are noisy or do not faithfully represent the task of ACR.
To achieve diversity, the examples represent 199 repositories across nine of the most popular programming languages on GitHub.
The evaluation set consists of 900 real code reviews in total.

\textbf{3) Comprehensive Evaluation of LLMs.}
A large scale evaluation of state-of-the-art open source LLMs that have been trained on both code and natural language.
These models span across five different scales, ranging from 1B to 72B parameters.
This includes code focused models e.g., CodeLlama-70b-Instruct-hf, general purpose models e.g., Qwen2.5-72B-Instruct, and the latest reasoning models e.g., QwQ-32B.
In total, we include 72 LLMs developed by 18 different organisations.

\section{Background and Related Work}
Recently, LLMs have shown promise in various software engineering tasks involving natural language artifacts. 
However, these artifacts vary significantly in their linguistic nature and structure. Some tasks involve explicit, non-conversational language, such as bug reports~\cite{bugsjar} and GitHub issues~\cite{swebench}, which typically contain detailed specifications of defects or feature requests. 
Other tasks involve static monologues, like commit messages~\cite{commitgen}, code comments~\cite{codecomment}, and pull request descriptions~\cite{prgen}, which aim to clearly explain source code or code changes. 

In contrast, code reviews are unique as they represent routine conversations in highly collaborative scenarios.
As such, they are informal, free-flowing, and can lean on the interlocutor's shared technical knowledge, without being overly specific~\cite{evacrc}.  
Thus, interpreting code reviews requires a deep understanding of conversational language in a highly technical context, posing challenges for computational models. 
The automated code refinement task involves revising source code to address code review comments, which requires an understanding of both technical implications and the reviewer's unstated expectations. 
Such nuanced communication makes automated code refinement an ideal testbed for evaluating LLMs' ability to bridge both technical and conversational understanding in software development.

The automated code refinement task was typically framed as a sequence-to-sequence neural machine translation problem, where models ``translate'' pre-review code submissions into post-review code revisions that reflect the intent of the accompanying code review comments. 
Formally, this problem requires the following estimation:
\begin{equation}
P(H_{post}|H_{pre},R_{nl})
\end{equation}
where $H_{pre}$ denotes the submitted pre-review code hunk, $R_{nl}$ denotes the natural language code review comment, and $H_{post}$ denotes the expected post-review revision of that code hunk. The ``hunk'' refers to the code snippet within the file, where the code review comment was inlined. See Figure~\ref{fig:mcqa} for a concrete example. 

While prior work has applied various neural language models, such as recurrent neural networks~\cite{tufano2019} and transformers~\cite{tufano2022, autotransform}, the task remains a challenging problem, even for recent LLMs such as GPT-4~\cite{guo2024exploring, llamareviewer, tufano2024code}. 

Indeed, prior work has highlighted several limitations in the evaluation~\cite{guo2024exploring}. 
Traditional evaluation approaches have relied heavily on text matching metrics such as exact match and BLEU~\cite{tufano2024code,guo2024exploring}, which are either too strict or fail to provide meaningful feedback. 
The emergence of LLMs has introduced additional challenges, as they are trained on extensive code repositories, creating significant risks of training data contamination. 
While some researchers have attempted to address this by collecting code reviews that outpace training cutoff dates~\cite{guo2024exploring}, such approaches lack long-term sustainability as real code reviews take time to naturally occur. 
Furthermore, existing benchmarks have been constructed automatically through large-scale mining using general heuristics, as a result, significant proportions of noise have been reported~\cite{tufano2024code,Liu2025Noisy}, undermining the reliability of past results. 

Table~\ref{table:compare} summarises the limitations of existing ACR evaluation benchmarks, underscoring the need for a new evaluation approach and dataset to reliably assess the capabilities of state-of-the-art LLMs. 
Our proposed \textbf{CodeReviewQA} benchmark focuses on addressing these gaps.

\input{Tables/benchmark_compare}
\input{Figures/mcqa_example}

\section{CodeReviewQA: MCQA Probes}
\label{sec:code_review_comprehension_probes}
Effective ACR relies heavily on the ability to comprehend $R_{nl}$ under the context of $H_{pre}$.
Rather than focusing this task as a sequence-to-sequence translation problem like the prior works, we argue that the model must be able to: 1) reason about the type of change $R_{nl}$ is requesting and 2) identify the relevant lines of code in $H_{pre}$ that is the subject of the change; and 3)  formulate the required code changes from a wide action space of potential code edits that can be performed on $H_{pre}$, before generating the code revision $H_{post}$.
The inability to perform the final code generation step may be caused by any failure point amongst this multi-step reasoning process.
Additionally, any failure within the intermediary reasoning steps might be propagated from a failure in a prior reasoning step, which obfuscates the specific incompetencies of the model.

To assess the proficiency of LLMs in ACR, we propose three MCQA probes, each representing a specific intermediate reasoning step. 
See Figure~\ref{fig:mcqa} for an example of the MCQA structure. 
Below, we describe the construction of each MCQA probe.

\subsection{Change Type Recognition (CTR)}
This is a closed set intent classification task that probes the model's ability to infer the intended type of code change.
Specifically, given $H_{pre}$, the model must infer which general type of code change is being requested by $R_{nl}$.
Formally, this problem requires the following estimation:

\begin{equation}
P(C_{type^+}|H_{pre},R_{nl})
\end{equation}

where $C_{type^+} \hspace{-0.3em} \in \hspace{-0.3em} \{add, delete, modify\}$ denotes the correct code change type.
There are three general types.
Firstly, \textit{add} requests involve only adding new lines of code.
Secondly, \textit{delete} requests involve only deleting existing lines of code.
Lastly, \textit{modify} requests involve altering the existing code by both deleting existing segments and adding new ones.
The $C_{type^-}$ distractors are the remaining two incorrect code change types.

This preliminary understanding serves as crucial conditional information that refines the problem space, providing the correct $C_{type}$ context to subsequently locate where the code changes need to occur and identify what needs to be implemented.

\subsection{Change Localisation (CL)}
This is a coreference resolution task that probes the model's ability to locate where the intended code change is to occur.
Specifically, given $R_{nl}$, the model must locate the precise lines of code within $H_{pre}$ where the intended $C_{type}$ code change should be applied.
Formally, this problem requires the following estimation:

\begin{equation}
P(C_{loc^+}|H_{pre},R_{nl},C_{type})
\end{equation}

where $C_{loc^+}$ denotes the exact set of line numbers that is the target of the intended code change.
When $C_{type} \hspace{-0.3em} \in \hspace{-0.3em} \{delete, modify\}$, these are the exact lines of code that need to be deleted or modified. 
When $C_{type}=\{add\}$, these are the lines of code above where the new code needs to be added.
The $C_{loc^-}$ distractors are different sets of lines sampled from $H_{pre}$. 
We ensure $|C_{loc^-}| = |C_{loc^+}|$, such that set sizes do not reveal additional information.

As shown in Figure~\ref{fig:mcqa}, natural code review comments often do not directly specify the exact location of the intended code change, rather this is implicitly conveyed based on a shared understanding between the reviewer and code author.
Thus, the model must possess the ability to conduct anaphora resolution across modalities, between anaphors in $R_{nl}$ and antecedents in $H_{pre}$.
Inferring the incorrect $C_{loc}$, would subsequently hinder the model's ability to identify the $H_{post}$ that accurately reflects the intended code change.

\subsection{Solution Identification (SI)}
This task probes the model's ability to both conduct open intent extraction from $R_{nl}$ and identify the $H_{post}$ that accurately reflects that intent. Given $R_{nl}$, the model must identify the correct $H_{post}$ that reflects the intended $C_{type}$ change on $C_{loc}$ in $H_{pre}$. The intuition behind this task design is that if a model is able to generate a correct $H_{post+}$ revision, it should at least be able to identify that exact $H_{post+}$ solution amongst a solution space with incorrect $H_{post^-}$ alternatives.   Formally, this problem requires the following estimation:
\begin{equation}
P(H_{post+}|H_{pre},R_{nl},C_{type},C_{loc})
\end{equation}
where $H_{post+}$ denotes the diff of the ground truth post-review code revision. We only include cases where $C_{type} \in \{add,modify\}$, as $\{delete\}$ cases merely delete $C_{loc}$ located in the previous task.

\subsection{Variation of Distractor Difficulty}
\label{sec:distractor}
The MCQA format allows flexibility in varying the difficulties of the distractors (i.e., the incorrect answer options).
This not only allows us to stress test the models' level of understanding, but also enables the ability to evolve the benchmark against performance saturation.
We specify the process of generating easy and hard distractors for \textit{Change Localisation} and \textit{Solution Identification}, as these tasks allow for variation in the solutions.

\textbf{\textit{Change Localisation Distractors}}. 
We vary the difficulty based on the degree of overlap between the sets of the provided $C_{loc}$ options.
For the easy distractors, we sample $C_{loc^-}$ distractors from $H_{pre}$, such that the \textit{Jaccard Similarity} between all answer options are as low as possible.
This ensures that all answers are easy to distinguish from each other and the ground truth is more obvious to locate.
For the hard distractors, we sample $C_{loc^-}$ distractors, such that the \textit{Jaccard Similarity} between all answer options are as high as possible.
This ensures that all answers are hard to discern from the ground truth, and requires the model to locate every exact line of the intended code change.

\textbf{\textit{Solution Identification Distractors}}. 
To create distractor options $H_{post-}$, we generate modified versions of $H_{post+}$ by perturbing code elements in $C_{loc}$, ensuring the intended code change is no longer correctly implemented. 
To create plausible but incorrect distractors that imitate possible mistakes that the models would make, we use a surrogate LLM~\footnote{We use a competitive surrogate LLM of code  (\href{https://huggingface.co/mistralai/Codestral-22B-v0.1}{Codestral-22B-v0.1}) with a temperature of 3.5 to encourage diversity.} to 1) identify the code element with the highest average surprisal in $H_{post+}$, 2) mask it, and 3) retroactively fill the masks with diverse candidates. 
We keep candidates that are not equivalent to $H_{post+}$ as valid $H_{post-}$ distractor candidates. All generated distractors are manually verified for semantic in-equivalence to the ground truth. 
The algorithm for constructing $H_{post-}$ distractors is illustrated in Algorithm~\ref{alg:mutant} in the Appendix.

We vary the difficulty based on the degree of semantic similarity between the $H_{post^-}$ distractors and the $H_{post^+}$ ground truth.
For the easy distractors, we retain the $H_{post^-}$ distractors which yield the lowest \textit{cosine similarity} against $H_{post^+}$ in the embedding space of the surrogate model.
This ensures that each $H_{post^-}$ is substantially different to $H_{post^+}$, such that it is easy to discern.
For the hard distractors, we retain the $H_{post^-}$ distractors which yield the highest \textit{cosine similarity} against $H_{post^+}$.
This ensures that each $H_{post^-}$ is only marginally different from $H_{post^+}$, such that it is hard to discern.
See Figure~\ref{fig:difficulty_variation} in the Appendix for examples of variation in difficulty.

\section{Dataset Preparation}

\textbf{Data Source.}
We built our benchmark based on the most recently published automated code refinement dataset~\cite{guo2024exploring}.
This multilingual dataset was constructed from code reviews that occurred after January 1, 2022. 
To ensure that we have a sizable amount of clean data for each of the programming languages in our benchmark, we only include the nine most popular programming languages on GitHub i.e. C, C++, C\#, Go, Java, Javascript, PHP, Python, Ruby.
These 9,367 examples were mined from 259 repositories, filtered from a list of the most starred GitHub projects.

\textbf{Data Sampling.}
To ensure diversity and quality in our benchmark, we conducted stratified sampling~\cite{baltes2022sampling} across all nine programming languages in the dataset, and discarded any examples that were noisy or unfaithfully represented the task of code refinement.
For each of the nine languages, we sampled until there were 100 clean examples each, resulting in 900 total examples in our benchmark.
Within each language partition, we also conducted stratified sampling across projects to maintain diversity.
This mitigated bias towards the code reviews of any specific project, the nature of which are influenced by their particular software development tools~\cite{reusability}, processes~\cite{dompractices} and issues~\cite{domainmatters}.

\textbf{Data Curation.}
We discard examples that were noisy or unfaithfully represented the task of code refinement.
The noisy examples refer to code review comments that are unclear, ignored, not asking for a code change, or linking to wrong code hunks. 
See Appendix~\ref{app:data_quality_issue_details} for a detailed explanation of these noise types.
These kind of review comments were reported as critical quality issues with existing code review datasets by prior work~\cite{tufano2024code}.
Unfaithful examples refer to the scenarios that do not faithfully represent the ACR task i.e., reviews directly including the intended $H_{post}$ revision implementation, reviews regarding simple code formatting, reviews that are not self-contained~\cite{tufano2024code,Lin2024}.
Instead, examples in the benchmark should represent meaningful quality improving code reviews that are beyond the capacity of traditional rule-based software engineering tools.
See Appendix~\ref{app:unfaithful_example_details} for a detailed explanation of these unfaithful examples.

To discard noisy and unfaithful examples, we first applied heuristic filters as detailed in Appendix~\ref{app:heuristic_filtering}, before manual verification.
This resulted in 3,761 out of 9,367 examples being discarded from the source dataset.
The manual discarding was conducted by the first and third authors, who are currently pursuing computer science PhDs focused on AI for software engineering.
In the end, both annotators independently annotated 3k examples, and resolved all conflicts together across 46 rounds.
The $\mu$ and $\sigma$ of the Cohen's Kappa were 0.89 and 0.11, respectively.
For C, JavaScript and Ruby, less than 100 clean examples could be obtained from the source dataset, thus, the remaining examples were sampled from code reviews conducted in 2021~\cite{codereviewer}.
The overall retention rate was 13\%, highlighting critical quality issues in the source datasets, necessitating manual curation of ACR benchmarks for accurate and reliable evaluation.
The final benchmark includes 199 of the original 259 GitHub repositories.
Table~\ref{table:statistic} in the Appendix shows benchmark statistics.

\section{Experimental Setup}
\subsection{MCQA Setup}
To support the MCQA probes in \textbf{CodeReviewQA}, we detail our prompt design, answer extraction, and evaluation framework for invariance testing.

\textbf{Prompt.} We use multiple-choice prompting that takes an input containing three components: task definition, question, and options. The task definition specifies the broad purpose (e.g., ``tests code review comprehension''). The question section presents the code review scenario within a template that includes programming language markers, $H_{pre}$, and $R_{nl}$. 
Finally, the options section lists multiple choice answers labelled alphabetically (A, B, C, D), with explicit instructions to respond with only the letter symbol. 
This prompt structure is used across all tasks, varying only the specific task parameters and answer options.
See Figure~\ref{fig:prompt_templates} in the Appendix for all prompt templates used.

\textbf{Answer Extraction.} We use multiple choice prompting with a max output length of one, where the symbol token $\in \{A, B, C, D\}$ with the highest log probability is considered as the selected answer.
This style of prompting avoids the conflation of likelihood of sequence and likelihood of answer, eliminates the need for normalisation and allows for direct comparison between answers~\cite{MCSB}.
Our implementation uses the vLLM inference framework~\cite{vLLM} with guided decoding targeting the option symbols.

\textbf{Invariant Test and Evaluation.}
To reduce the likelihood of random correct guesses, for each question, we exposed the models to every order combination of the answer options.
This resulted in $N!$ runs per question, where N is the number of answer options provided. 
Thus, the likelihood of guessing the correct answer for all combinations of a question is merely $(\frac{1}{N})^{N!}$.
With this, we assess the models' invariability in selecting the correct answer, regardless of the position of that answer. 
To be counted as correctly answering that question, the models must select the correct answer for all $N!$ runs, which is a more reliable indicator of the models' understanding~\cite{beyondanswers}.

\subsection{Model Selection Criteria}
We list the criteria that determines whether a LLM is appropriate for this benchmark. 

\textit{\textbf{MCQA Proficiency.}}
The LLM must have achieved state-of-the-art results in MCQA style benchmarks e.g., MMLU~\cite{MMLU}.
This accounts for format as a confounding factor.

\textit{\textbf{MCSB Proficiency.}}
The LLM must demonstrate proficiency in multiple choice symbol binding (MCSB;~\citet{MCSB}).
This ensures that the answer extraction method is not a confounding factor.
We report the Proportion of Plurarity Agreement (PPA), which measures the degree of order invariance in selecting the symbol of the plurarity answer.
Formally, PPA is calculated as the average of $\frac{k}{N!}$ over a dataset, where $k$ is the number of times the plurarity answer's symbol yielded the highest log probability for a given question and $N!$ is the aforementioned number of order combinations for $N$ answer options.
MCSB proficiency is demonstrated when a PPA significantly higher than the random baseline of $\frac{1}{N}$ is achieved.

\textit{\textbf{Coding Proficiency.}}
In addition to understanding the natural language in $R_{nl}$, the model must also be able to understand the code in $H_{pre}$ and $H_{post}$. 
Therefore, the LLM must have demonstrated proficiency in coding related benchmarks e.g., HumanEval~\cite{humaneval} and MBPP~\cite{MBPP}.

In total, we selected 72 state-of-the-art open source LLMs, that have satisfied the three criteria. 
The included models are considered state-of-the-art as of March, 2025.
See Table~\ref{table:model} in Appendix~\ref{app:experimental_results_72models} for descriptions of all 72 models. 
The models are grouped into five scales based on their model parameters: $\leq$3B, $\leq$9B, $\leq$16B, $\leq$34B, and $\leq$72B. 
We selected models with $\leq$72B parameters, as it is the largest non-quantised model class that we can support locally to extract answer probabilities.

\section{Results Overview}
To compare how models of different scales perform on \textbf{CodeReviewQA}, we conducted experiments using all 72 models. 
Due to space limitations, complete results on ACR and all three MCQA probes are presented in Appendix~\ref{app:experimental_results_72models}.
In Table~\ref{table:top_models}, we compare the MCQA probe results of the top-2 models in terms of ACR from each scale class. 

\textbf{ACR vs MCQA Probes.} 
Table~\ref{table:top_models} (column ACR) shows that Llama-3.1-70B-Instruct achieved the highest exact match rate of 50.3\%.
As expected, we find that larger LLMs tend to achieve higher exact match rates on average in ACR.
However, the performance improvements appear to be diminishing when scaling past the $\leq$16B class of models.
Specifically, we find that Llama-3.1-70B-Instruct only achieves a 3.7\% increase over Qwen2.5-Coder-14B-Instruct, despite wielding a 5-fold increase in parameter count. 
Interestingly, we find that ACR performances are not always congruent with how models rank in MCQA probes.
For example, Qwen2.5-72B-Instruct achieves an exact match rate approximately 2\% lower than the top performer, but performs vastly better (> 10\% increase) in terms of invariant accuracy in both CTR and SI.
This type of inconsistency also occurs throughout the smaller scales.
For example, Qwen2.5-Coder-14B-Instruct, gemma-2-27b-it, and Mistral-Small-Instruct-2409 achieve comparable exact match rates ($\leq$ 1\% difference) in ACR. 
However, their performances on CL and SI probes are substantially different.
These additional insights highlight the benefits of intermediate feedback, beyond simple text matching evaluations on final code revision outputs. 
Below, we discuss results on the three MCQA probing tasks.

\input{Tables/Top_Models}
\textbf{CTR Results.}
Table~\ref{table:top_models} (column CTR) shows that most of the $\leq3$B models were already competent in this task, with Llama-3.2-3B-Instruct achieving 78.8\% invariant accuracy.
Despite the promising results of small models, this ability plateaus as we scale model size.
In fact, Qwen2.5-72B-Instruct only achieved a 1\% improvement, despite having 24 times the amount of parameters.
Interestingly, Llama-3.1-70B-Instruct even regresses in this capability, achieving the worst invariant accuracy across the top performing models.
As knowledge of the general change type sets the entire context for deciding where and how to revise the code, the inability to improve on CTR presents a fundamental limit to downstream ACR performance.

\textbf{CL Results.}
Table~\ref{table:top_models} (columns CL$_\text{E}$ and CL$_\text{H}$) show that change localisation tends to be the most difficult comprehension task in the benchmark, where proficiency depends more heavily on scale.
Table~\ref{table:cl_easy} in Appendix~\ref{app:experimental_results_72models} shows the full results for CL$_\text{E}$, where most of the $\leq3$B models achieved invariant accuracies between 0\%-3\%.
The only exceptions were Qwen2.5-3B-Instruct and Phi-3-mini-128k-instruct, which could achieve 39.3\% and 34.1\%, respectively.
The discrepancy between ACR and CL results across the under parameterised models suggest a reliance on surface-level pattern recognition for ACR, that is not grounded in a robust understanding of the code review.
In contrast, we find that many models from the $\leq34$B and $\leq72$B classes could achieve invariant accuracies of more than 70\% for CL$_\text{E}$ and more than 60\% for CL$_\text{H}$.
Nevertheless, proficiency in identifying exactly where the code revisions need to occur remains the largest challenge. 

\textbf{SI Results.}
Table~\ref{table:top_models} (columns SI$_\text{E}$ and SI$_\text{H}$) show that proficiency in solution identification also strengthens with size.
Despite this, Table~\ref{table:si_easy} in Appendix~\ref{app:experimental_results_72models} shows a few anomalies that outperform their scale class average by a large margin.
For example, Phi-3-mini-128k-instruct can achieve an invariant accuracy of 58.2\% for SI$_\text{E}$, whilst the majority of $\leq$3B models achieve less than 13\%.
In contrast, many models from the $\leq$72B class can achieve more than 80\% for SI$_\text{E}$ and more than 70\% for SI$_\text{H}$.
Most notably, Qwen2.5-72B-Instruct, could achieve a near perfect score of 97.1\% for SI$_\text{E}$, and 90.9\% for SI$_\text{H}$, despite previously achieving underwhelming results for change localisation.
These results suggest that many larger models can identify the correct code revision at least under a drastically reduced solution space with semantically ambiguous distractors.

\section{Insights from Probing ACR Failures}

\textbf{What does MCQA probes reveal about model failures in ACR?}
We use our MCQA probes to investigate ACR failure cases of the top performing models in Table~\ref{table:top_models}. 
We define a failure as the case when the model does not generate a $H_{post}$ revision that exactly matches the ground truth.
For each failure case, we examine whether the model also fails on the MCQA probes.
Through this analysis, we aim to identify specific weaknesses in the models' code review comprehension capabilities that may contribute to their ACR failures.

The results are presented in Table~\ref{table:probing}, including the percentage of failure on each of the MCQA probes,~\footnote{For CL and SI, a failure in either difficulties is counted.} and the overall failure on at least one probe ($\geq 1$ Probe). 
We find that all five models failed at least one MCQA probe for 76.5\%-99.8\% of the failure cases, indicating that the model struggled at the code review comprehension stage.
CTR failures are seldom associated with the ACR failures, as this capability only accounts for 22.2\%-37.4\% of cases.
In contrast, CL failures are often associated with ACR failures, particularly for the three smaller models, where 74.0\%-99.4\% of the cases overlap with ACR failures.
SI failures account for 95.8\% of cases for Qwen2.5-Coder-3B-Instruct.
Thus, for the smallest model, most non-exact matches are associated with failures in both CL and SI.
For both Qwen2.5-Coder-7B-Instruct and Qwen2.5-Coder-14B-Instruct, CL failure is the major factor associated with ACR failure.
For the larger models, gemma-2-27b-it and Llama-3.1-70B-Instruct, the probe failures are more evenly distributed, varying based on the model's specific incompetencies.

On the other hand, we also analysed the successful cases where the model achieves an exact match.
Intuitively, if a model can achieve exact match on an example, it should be able to fully comprehend the code review, thus correctly answering all probes.
However, the models could not accurately answer all probes for 49.0\%-99.6\% of their successful cases.
This trend shows a strict inverse relationship with model size, where Llama-3.1-70B-Instruct and Qwen2.5-Coder-3B-Instruct could not consistently complete all of the probes for 49.0\% and 99.6\% of their successful cases, respectively.
The ability to complete these ACR examples verbatim without a prerequisite understanding of the intent behind the code reviews allude to prior exposure and rote memorisation of these examples.

\input{Tables/probing}

\section{Evaluating Data Contamination}
\label{sec:evaluating_data_contamination}

\textbf{To what extent is \textbf{CodeReviewQA} resistant to data contamination?}
We utilise two canonical metrics for measuring data contamination, perplexity~\cite{Jelinek1977PerplexityaMO} and n-gram accuracy~\cite{xu_ngram}.
Perplexity is an information-theoretic metric, which quantifies the uncertainty of a language model in a token sequence~\cite{Jelinek1977PerplexityaMO}, which can be formulated as:
\begin{equation}
\text{PPL}(\textbf{X}) = \exp(-\frac{1}{t}\sum^{t}_{t=0}\log p_{\theta}(x_{i}|x_{<i}))
\end{equation}

where \textbf{X} = [$x_{0}, x_{1}, ... , x_{t}$] denotes a tokenised sequence. 
In our case, the sequence is a concatenation of both the prompt (including $H_{pre}$, $R_{nl}$) and the solution.
For ACR, the solution is $H_{post}$, and for the MCQA probes, it is the answer options.
A low perplexity score indicates high confidence, whilst a high perplexity score indicates low confidence.
Unusually low perplexity scores may indicate data contamination.
N-gram accuracy measures the model's ability to predict random n-gram sequences from $K$ starting points that are uniformly sampled from an example~\cite{xu_ngram}, i.e., the aforementioned sequence $X$.
It is calculated by the following equation:

\begin{equation}
\text{NG}(\textbf{X}) = \frac{1}{\eta \cdot K} \sum^{\eta}_{i=0} \sum^{K}_{j=0} I(X_{s_{j}:s_{j}+n}, \hat{X}_{s_{j}:s_{j}+n})
\end{equation}

where $\eta$ denotes the corpus size, $i$ denotes the $i_{\text{th}}$ sequence in the corpus, $s_{j}$ denotes the index of the $j^{th}$ starting point, $X_{s_{j}:s_{j}+n}$ denotes the ground truth n-gram to be predicted and $I$ denotes an indicator function that applies exact match.
Unusually high n-gram accuracies may indicate data contamination.
Following prior work~\cite{xu_ngram}, we set $K$ = 5 and $n=5$ to measure 5-gram accuracy.

For this experiment, we use the largest and newest models that we can support from the most popular model families, as they are most likely to exhibit memorisation~\cite{kiyomaru_memorisation}.
We use base versions of models as instruction-tuning optimises for responses to prompts rather than completing sequences verbatim.
To test the effectiveness of MCQA reformulation in mitigating data contamination, we compare our benchmark in MCQA probe form with the original ACR form, as well as the most widely used ACR benchmarks, i.e. CodeReviewer~\cite{codereviewer} and CodeReview-New~\cite{guo2024exploring}.

\input{Tables/data_contamination}

Table~\ref{table:data_contamination} shows the perplexity and 5-gram accuracies on the three benchmarks, based on two popular base models Llama-3.1-70B and Qwen2.5-72B.
We find that perplexity on the older CodeReviewer benchmark is far lower than on CodeReview-New, yet the 5-gram accuracies are also lower.
A likely explanation is that older code reviews may have been extensively included in the models' training phase and therefore reflect the predominant patterns in their learned distribution, however, since they may not have been included in the latter stages, there is less verbatim memorisation of the examples~\cite{kiyomaru_memorisation}.
In contrast, the newer code reviews may represent a distribution shift, yet is more likely to be included in the latter stages of training, thus concurrently increasing both perplexity and verbatim memorisation at the same time.

We find that \textbf{CodeReviewQA} in the original ACR format yields similar results to CodeReview-New, which is within expectation as one is simply a curated subset of the other.
However, when reformulating into the MCQA probe format, our benchmark yields significantly higher perplexity than all past benchmarks with lower 5-gram accuracies, despite using the same examples.
Therefore, we find that MCQA reformulation with synthetic questions and answers does mitigate the effects of data contamination, allowing for the reuse of code reviews that may have been previously included in the training corpus. 
Coinciding with our experimental results, models that perform well on ACR with only memorisation can be exposed when evaluated with MCQA probes on the same examples.

\section{Conclusion}

In this study, we focus on evaluating recent large language models' capabilities in automated code refinement, a challenging task that requires an understanding of the intended code revisions behind natural language code reviews, before subsequently performing them. 
We addressed two key limitations in existing work, the inability of text matching metrics to provide fine-grained feedback on intermediate model failures and the potential for benchmark contamination. 
To this end, we propose \textbf{CodeReviewQA}, which consists of 900 manually curated high-quality code review examples.
We reformulated the generative task of automated code refinement into three intermediate reasoning probes, which are presented in the multiple choice question and answering format. 
Our experimental results across 72 state-of-the-art large language models revealed capability differences that traditional evaluation metrics failed to capture. 
Additionally, our evaluation of data contamination demonstrated that task reformulation effectively mitigates these concerns, exposing cases of memorisation without comprehension.

\section{Limitations}
Whilst \textbf{CodeReviewQA} advances the evaluation of automated code refinement, it still faces limitations.

\textbf{Size of dataset.} 
Our benchmark has a relatively modest size due to the difficulty of scaling rigorous manual verification.
Despite this, the benchmark was designed to be diverse and comprehensive through stratified sampling, covering real-world code reviews from 199 different GitHub projects in nine of the most popular programming languages.

\textbf{Construction of distractors.} 
The change localisation task focused on line level localisation rather than a more fine-grained level (e.g., token level).
However, our findings show that many models struggle with identifying the location of changes even at the line level. 
Future work can further explore approaches to automatically construct and evaluate localisation at the token level. 
The solution identification task relies on a competitive surrogate LLM for constructing challenging distractors.
Whilst Codestral-22B-v0.1 is considered competitive at the time of this writing, benchmark saturation remains inevitable as more powerful LLMs are developed.
Nonetheless, the benchmark can be evolved by swapping to a more up-to-date surrogate model in the future, thus increasing the difficulty and relevance of the distractors.

\textbf{Interaction among capabilities.} 
As we were interested in analysing the successive code review comprehension capabilities in isolation, each probe was designed to be independent of each other by including the preceding probe's ground truth as conditional information.
We did not investigate the causal relationships between the capabilities tested in the probes, meaning that failure in one probe does not predict performance on another. 
However, our experimental results demonstrate that analysing disentangled capabilities alongside the automated code refinement task provides more interpretable insights into model weaknesses.

\textbf{Diversity of prompts.} 
We used the same prompt and hyperparameters for each task to maintain consistency and comparability across models. 
Prompt variation might impact model performance differently.
However, our main focus was not to find the optimal prompt for each model, but to gauge their systematic differences across different capabilities required for automated code refinement.

\textbf{Semantic level data contamination.}
Our MCQA reformulation process only mitigates the effects of surface level data contamination i.e., when examples are presented as a next token prediction task in their original form.
Given that the actual content of the examples have not been altered, there still exists the risk of data contamination at the semantic level, where models can leverage the learned meanings conveyed in the code reviews that they have been exposed to.
To the best of our knowledge, there are no viable solutions that can completely address this concern, apart from collecting new code reviews or using completely private projects, which poses concerns of feasibility and generalisability.
Nevertheless, our benchmark evaluates knowledge invariance in comprehending and addressing code reviews.
Assessing this type of generalisation of the contaminated examples can still be considered a meaningful measurement of learned capabilities, beyond simple memorisation.

\bibliography{custom}

\appendix
\section{Data Quality Issue Details}
\label{app:data_quality_issue_details}
We explain the four types of data quality issues found in code review datasets.
Firstly, \textit{unclear comments} are review comments where even humans cannot comprehend the intended change.
Secondly, \textit{no change asked} refers to review comments that are not actionable.
Thirdly, \textit{ignored comment} are examples where the developer ignores the review comment, resulting in a post-review code revision $H_{post}$ that does not reflect the intended code change.
Lastly, \textit{wrong linking} refers to data mining issues, where the review comment is not related to the paired pre-review code submission $H_{pre}$.

\section{Unfaithful Example Details}
\label{app:unfaithful_example_details}
We explain the three types of unfaithful examples found in code review datasets.
Firstly, some code reviews directly include the \textit{intended code revision implementation}. These cases can be resolved by directly copy and pasting from the review itself, which does not assess natural language comprehension.
Secondly, \textit{code formatting} related examples can already be resolved by linters and therefore are not useful to learn.
These examples also fail to assess the models' ability in handling challenging and meaningful code reviews.
Thirdly, code reviews that are \textit{not self-contained} require information beyond the provided code hunk $H_{pre}$ to understand, thus, it is impossible for the model (or even a human) to intuit the intended code change.

\section{Heuristic Filtering Details}
\label{app:heuristic_filtering}
We conducted keyword-based filtering to automatically discard examples that clearly violate the data quality and faithfulness issues mentioned above.
With regards to \textit{unclear comments}, we discarded reviews with less than 10 characters, since they are likely to be too short to convey a code change requirement.
With regards to code reviews that already contain the \textit{intended code revision implementation},
we discarded reviews that included the "```" GitHub code block indicator.
With regards to code reviews that are only demanding \textit{code formatting} changes, we removed reviews that mention "indentation", "spacing" and "lint".
With regards to reviews that are \textit{not self-contained}, we removed reviews that mention "revert", "as above" and "ditto".
The purpose of this step was to reduce human workload in the proceeding manual discarding process.

\section{Descriptive Statistics}

\label{app:benchmark_statistic}
We explain the descriptive statistics used in this study for describing the benchmark.

\textbf{Comment length} is measured by the number of whitespace separated words in the code review comment.
Longer comments may contain more complex requirements, explanations or other discussions.
Table~\ref{table:statistic} shows that the average comment length is 18 words.
See Figure~\ref{fig:comment_length} for examples of different comment lengths.

\textbf{Code edit distance} represents the size of the code change between $H_{pre}$ and $H_{post}$.
Given that some examples only involve changes in the inlined code comment, we use the more general Levenshtein distance~\cite{levenshtein1966binary} to measure the number of character edits between the two versions of code.
Table~\ref{table:statistic} shows that the average code edit distance is 56 characters.
See Figure~\ref{fig:code_edit_distance} for examples of different code edit distances.

\textbf{Change locations} is the number of lines involved in the change, as discussed in the task of change localisation.
Table~\ref{table:statistic} shows the change location statistics.
For the 35 (4\% of total) code reviews that request to \textit{add} code, the average number of change locations is 1 line.
For the 144 (16\% of total) code reviews that request to \textit{delete} code, the average number of change locations is 3 lines.
For the 721 (80\% of total) code reviews that request to \textit{modify} code, the average number of change locations is 2 lines.

\input{Tables/statistics}

\textbf{Code element ratio} is the proportion of tokens in the code review comment that are code elements.
It is calculated as $\frac{\text{Code elements}}{\text{Comment length}}$.
Reviewers may use code elements in conjunction with natural language to describe the intended code change. Comments with a higher proportion of code tokens may be more explicit in their specification of the requirements~\cite{cr_usefulness}.
Table~\ref{table:statistic} shows that the average code element ratio is 0.09.
See Figure~\ref{fig:code_element_ratio} for examples of different code element ratios.

\textbf{Specification ratio} is the code edit distance of the change divided by the length of its respective code review comment.
It is calculated as $\frac{\text{Code edit distance}}{\text{Comment length}}$, the number of character edits with respect to each word in the comment.
Since code review comments may be under-specified and implicit, we use specification ratio as a heuristic metric that incorporates this notion.
Intuitively, examples with larger code edit distances and shorter comment lengths i.e. larger specification ratios, may be under-specified in its description of the required code change.
Table~\ref{table:statistic} shows that the average specification ratio is 4.88.
See Figure~\ref{fig:specification_ratio} for examples of different specification ratios.

\input{Figures/prompt_templates}
\input{Figures/fewshot_CTR}
\input{Figures/fewshot_CL}
\input{Figures/fewshot_SI}
\input{Figures/difficulty_variation}
\input{Figures/comment_length}

\input{Figures/code_edit_distance}
\input{Figures/code_element_ratio}
\input{Figures/spec_ratio}

\section{Implementation Details}
The experiments were carried out on a single node consisting of 64 cores (Intel(R) Xeon(R) Platinum 8462Y+ @ 2.80GHz), 928GBs of RAM and four GPUs (NVIDIA H100-80GB SXM5).
For efficient inference, we utilised the default vLLM implementation with Hugging Face models.~\footnote{~\href{https://huggingface.co/docs/hub/en/models-the-hub}{https://huggingface.co/docs/hub/en/models-the-hub}}
For all tasks, we set the temperature to zero for greedy decoding.

\section{Experimental Results on 72 models}
\label{app:experimental_results_72models}
This section shows our complete results.
The details of all evaluated LLMs are presented in Table~\ref{table:model}, and their achieved exact match rate on ACR in Table~\ref{table:acr}.
For the MCQA probes, we report both their proportion of plurarity agreement and invariant accuracy. 
If the achieved proportion is far higher than the stated random probability on any probe, the model is considered to have multiple choice symbol binding proficiency. 
Table~\ref{table:ctr} shows the full results on CTR.
Table~\ref{table:cl_easy} and Table~\ref{table:cl_hard} shows full results on the easy and hard variations of CL, respectively.
Table~\ref{table:si_easy} and Table~\ref{table:si_hard} shows full results on the easy and hard variations of SI, respectively.

\section{Advanced Prompting}
Since our main results were based on zero-shot prompting, we further investigated the effects of advanced prompting i.e., few-shot, chain-of-thought, on MCQA probe performance.
Specifically, we were interested if advanced prompting could increase the performance of the top performing model i.e., Llama-3.1-70B-Instruct.
For few-shot prompting, we considered both one-shot and two-shot scenarios with real examples held out from the benchmark. 
For the few-shot prompt templates, see Figure~\ref{fig:fewshot_ctr} for CTR, Figure~\ref{fig:fewshot_cl} for CL and Figure~\ref{fig:fewshot_si} for SI.
For chain-of-thought, we only considered zero-shot due to the lack of real reasoning traces to use as examples.
This was invoked by appending "\textit{Let's think step by step}" to the end of the zero-shot prompts and instructing the model to finalise their answer with "\textit{The final answer is}".
\input{Tables/advanced_prompting}

The results are shown in Table~\ref{table:advanced_prompting}.
Overall, we find that advanced prompting does not outperform the zero-shot strategy.
For CTR, we find that zero-shot performed similar to chain-of-thought, where both accuracies were approximately 68\%.
In contrast, the few-shot methods could achieve 74.1-76.9\%, where two-shot prompting was the top performing prompt strategy.
In terms of CL easy, we find that one-shot and chain-of-thought were the worst performers, achieving 68.6\% and 65.8\%, respectively.
In contrast, zero-shot and two-shot performed similarly, achieving 74.7\% and 75.8\%, respectively.
For the remaining MCQA probes, all advanced prompting techniques underperformed against the zero-shot strategy, especially for SI.

Whilst chain-of-thought consistently degraded the model's performance on the MCQA probes, few-shot prompting showed promise.
Specifically, two-shot prompting was the top performing strategy in two of the five probes, without optimising for the best examples.
We did not optimise this selection due to the lack of clean code review examples.
As a result, we only used two held out examples collected during the manual curation process.
Future researchers can investigate the effects of advanced example selection e.g., retrieval-augmented generation, for few-shot prompting strategies.

\input{Figures/si_distract_alg}
\input{Tables/models}
\input{Tables/ACR}
\input{Tables/CTR}
\input{Tables/CL_easy}
\input{Tables/CL_hard}
\input{Tables/SI_easy}
\input{Tables/SI_hard}

\end{document}

%% file: Tables/benchmark_compare.tex
\begin{table}[t]
\resizebox{\columnwidth}{!}{
\begin{tabular}{llclccc}
 \toprule
\textbf{Code Review Benchmark} & \textbf{Size} & \textbf{\#Lang} & \textbf{Metric} & \textbf{DC} & \textbf{MV} & \textbf{VD} \\ \midrule
Tufano 2021~\cite{tufano2021}  & 1.7k & 1 &  Text Match &  \faTimes &  \faTimes &  \faTimes \\
T5CR~\cite{tufano2022} & 16.8k & 1 &  Text Match &  \faTimes &  \faTimes &  \faTimes \\
CodeReviewer~\cite{codereviewer} & 13.1k & 9 &  Text Match &  \faTimes &  \faTimes &  \faTimes \\
CodeReview-New~\cite{guo2024exploring}  & 14.6k & 16 &  Text Match &  \faTimes &  \faTimes &  \faTimes \\ \hline
\textbf{CodeReviewQA (Ours)} & 900 & 9 & \parbox{1.7cm}{\vspace{0.1cm} Text Match \\ \& Probe \vspace{0.1cm}} & \faCheck & \faCheck & \faCheck \\ 
\bottomrule
\multicolumn{7}{l}{\textbf{DC:} Addresses Data Contamination, \textbf{MV:} Manual Verification, \textbf{VD:} Varied Difficulty}
\end{tabular}
}
\caption{Benchmarks for ACR.}
\label{table:compare}
\end{table}

%% file: Figures/mcqa_example.tex
\begin{figure}[!t]
\begin{tcolorbox}[colback=gray!20, colframe=gray, 
left=0pt, right=0pt, arc=10pt, width=\columnwidth, boxrule=1pt]
\fontsize{7}{5}\selectfont 
\textcolor{purple}{\textbf{Pre-Review Code Submission ($H_{pre}$):}}

\vspace{1mm}
\begin{lstlisting}
    from hypothesistooling.projects.hypothesispython import PYTHON_SRC
    from hypothesistooling.scripts import pip_tool, tool_path

    PYTHON_VERSIONS = [f"3.{v}" for v in range(7, 11)]

    def test_mypy_passes_on_hypothesis():
\end{lstlisting}
\vspace{1mm}

\textcolor{purple}{\textbf{Code Review ($R_{nl}$)}:} 
I think I'd prefer to write these out as literals, unless we can pull them out of the autoupdated CI config? 
Just thinking about how they'll stay up to date. 
I think we can also test against 3.11?

\textcolor{gray}{\textbf{ ------------------------------------------------------------------------------------------ }}

\vspace{1mm}
\textcolor{purple}{\textbf{What type of change is the code review asking for?}}

A. Only add new lines of code

B. Only delete existing lines of code

\textcolor{dkgreen}{C. Modify the code $\checkmark$}

\textcolor{gray}{\textbf{ ------------------------------------------------------------------------------------------ }}

\vspace{1mm}
\textcolor{purple}{\textbf{Which line numbers is the code review asking to modify code?}}

A. line number 1 $\phantom{\checkmark}$ B. line number 2 $\phantom{\checkmark}$  

\vspace{1mm}
\textcolor{dkgreen}{C. line number 4 $\checkmark$} D. line number 6

\textcolor{gray}{\textbf{ ------------------------------------------------------------------------------------------ }}

\vspace{1mm}
\textcolor{purple}{\textbf{Which code revision is the code review asking for?}}

A. 
\begin{lstlisting}[numbers=none, escapeinside={(*@}{@*)}]
   (*@\textcolor{listinggray}{4}@*)  - PYTHON_VERSIONS = [f"3.{v}" for v in range(7, 11)]
   (*@\textcolor{listinggray}{4}@*)  + PYTHON_VERSIONS >= ["3.7", "3.8", "3.9", "3.10", "3.11"]
\end{lstlisting}
B. 
\begin{lstlisting}[numbers=none, escapeinside={(*@}{@*)}]
   (*@\textcolor{listinggray}{4}@*)  - PYTHON_VERSIONS = [f"3.{v}" for v in range(7, 11)]
   (*@\textcolor{listinggray}{4}@*)  + PYTHON_VERSIONS <= ["3.7", "3.8", "3.9", "3.10", "3.11"]
\end{lstlisting}
C. 
\begin{lstlisting}[numbers=none, escapeinside={(*@}{@*)}]
   (*@\textcolor{listinggray}{4}@*)  - PYTHON_VERSIONS = [f"3.{v}" for v in range(7, 11)]
   (*@\textcolor{listinggray}{4}@*)  + PYTHON_VERSIONS != ["3.7", "3.8", "3.9", "3.10", "3.11"]
\end{lstlisting}
\textcolor{dkgreen}{D. $\checkmark$}
\begin{lstlisting}[numbers=none, escapeinside={(*@}{@*)}]
   (*@\textcolor{listinggray}{4}@*)  - PYTHON_VERSIONS = [f"3.{v}" for v in range(7, 11)]
   (*@\textcolor{listinggray}{4}@*)  + PYTHON_VERSIONS = ["3.7", "3.8", "3.9", "3.10", "3.11"]
\end{lstlisting}

\textcolor{gray}{\textbf{ ------------------------------------------------------------------------------------------ }}

\vspace{1mm}
\textcolor{purple}{\textbf{Post-Review Code Revision ($H_{post}$)}:}
\begin{lstlisting}
   from hypothesistooling.projects.hypothesispython import PYTHON_SRC
   from hypothesistooling.scripts import pip_tool, tool_path
  
   PYTHON_VERSIONS = ["3.7", "3.8", "3.9", "3.10", "3.11"]
  
   def test_mypy_passes_on_hypothesis():
\end{lstlisting}

\end{tcolorbox}
\caption{The ACR task with intermediate reasoning steps presented as MCQA Probes.}
\label{fig:mcqa}
\end{figure}

%% file: Tables/Top_Models.tex
\begin{table}[!t]

\resizebox{\columnwidth}{!}{
\begin{tabular}{lcccccc}
\hlineB{2.7}
\textbf{Model} & \textbf{ACR} & \textbf{CTR} & \textbf{CL$_\text{E}$} & \textbf{CL$_\text{H}$} & \textbf{SI$_\text{E}$} & \textbf{SI$_\text{H}$} \\ \hline

\cellcolor{gray!20}Qwen2.5-Coder-3B-Instruct  &\cellcolor{gray!20}30.3 &\cellcolor{gray!20}77.7 &\cellcolor{gray!20}1.8 &\cellcolor{gray!20}1.8 &\cellcolor{gray!20}12.2 &\cellcolor{gray!20}8.0 \\

Llama-3.2-3B-Instruct	& 25.9	& 78.8	&0.8	&0.4	&9.9	&7.6 \\ 
\hline

\cellcolor{gray!20}Qwen2.5-Coder-7B-Instruct  & \cellcolor{gray!20}41.0 & \cellcolor{gray!20}78.6 &  \cellcolor{gray!20}13.8 & \cellcolor{gray!20}10.7 & \cellcolor{gray!20}67.6 & \cellcolor{gray!20}55.2 \\
gemma-2-9b-it	& 39.0	& 74.1	& 59.2	& 52.0	& 58.8	& 49.6 \\ 
\hline

\cellcolor{gray!20}Qwen2.5-Coder-14B-Instruct  &\cellcolor{gray!20}46.6 &\cellcolor{gray!20}73.9 &\cellcolor{gray!20}46.7 &\cellcolor{gray!20}37.3 &\cellcolor{gray!20}65.5 &\cellcolor{gray!20}56.2 \\

phi-4	&37.1 &76.6	&50.9 &44.8 &84.4 &77.5 \\ 
\hline

\cellcolor{gray!20}gemma-2-27b-it  & \cellcolor{gray!20}46.4 & \cellcolor{gray!20}74.0 &  \cellcolor{gray!20}70.1 & \cellcolor{gray!20}58.7 & \cellcolor{gray!20}76.2 & \cellcolor{gray!20}65.7 \\

Mistral-Small-Instruct-2409 	& 45.6	&76.7 	&38.4 	&31.6	&63.8	&60.1 \\ 
\hline 

\cellcolor{gray!20}Llama-3.1-70B-Instruct  & \cellcolor{gray!20}\textbf{50.3} & \cellcolor{gray!20}68.4 &  \cellcolor{gray!20}\textbf{74.7} & \cellcolor{gray!20}\textbf{69.0} & \cellcolor{gray!20}84.2 & \cellcolor{gray!20}76.7 \\

Qwen2.5-72B-Instruct	& 48.7	& \textbf{79.8}	& 64.2	& 58.3	& \textbf{97.1}	& \textbf{90.9}\\  
\hlineB{2.7}

\multicolumn{7}{l}{\textbf{ACR:} Automated Code Refinement, \textbf{CTR:} Change Type Recognition} \\
\multicolumn{7}{l}{\textbf{CL:} Change Localisation, \textbf{SI:} Solution Identification, \textbf{E:} Easy, \textbf{H:} Hard} \\
\end{tabular}
}

\caption{MCQA Probe Accuracy (\%) of top-2 models (per scale class) in terms of Exact Match Rate on ACR.}
\label{table:top_models}\
\end{table}

%% file: Tables/probing.tex







\begin{table}[!t]

\resizebox{\columnwidth}{!}{
\begin{tabular}{lcccc}
\hlineB{2.7}
 & \multicolumn{4}{c}{\textbf{\%Fail}}\\
\textbf{Model} & \textbf{$\geq1$ Probe} & \textbf{CTR} & \textbf{CL} & \textbf{SI}\\ \hline
  
\cellcolor{gray!20}Qwen2.5-Coder-3B-Instruct  & \cellcolor{gray!20}\textbf{99.8} & \cellcolor{gray!20}23.4 &  \cellcolor{gray!20}\textbf{99.4} & \cellcolor{gray!20}\textbf{95.8}  \\

Qwen2.5-Coder-7B-Instruct   & 96.6 & 22.2 &  93.0 & 55.0  \\

\cellcolor{gray!20}Qwen2.5-Coder-14B-Instruct   & \cellcolor{gray!20}87.3 & \cellcolor{gray!20} 29.3 &  \cellcolor{gray!20}74.0 & \cellcolor{gray!20}56.0  \\

gemma-2-27b-it  & 75.5   & 25.7 &  53.7 & 41.4  \\

\cellcolor{gray!20}Llama-3.1-70B-Instruct & \cellcolor{gray!20}76.5 & \cellcolor{gray!20}\textbf{37.4} &  \cellcolor{gray!20}46.3 & \cellcolor{gray!20}32.7 \\
\hlineB{2.7}

\multicolumn{5}{l}{\textbf{CTR:}  Change Type Recognition, \textbf{CL:} Change Localisation, } \\
\multicolumn{5}{l}{\textbf{$\geq1$ Probe:} Failed at least one probe, \textbf{SI:} Solution Identification} \\
\end{tabular}
}
\caption{MCQA Failure (\%) in Non-Exact Match cases of the top performing model (per scale class) in ACR.}
\label{table:probing}
\end{table}

%% file: Tables/data_contamination.tex
\begin{table}[!t]
\resizebox{\columnwidth}{!}{
\begin{tabular}{llcccc}

\hlineB{2.7}
\textbf{Benchmark} & \textbf{Format} & \multicolumn{2}{c}{\textbf{Llama-3.1-70B}} & \multicolumn{2}{c}{\textbf{Qwen2.5-72B}} \\ 
~ & ~ & \textbf{PPL} & \textbf{NG$_5$} & \textbf{PPL} & \textbf{NG$_5$} \\ \hline
\cellcolor{gray!20}CodeReviewer & \cellcolor{gray!20}ACR & \cellcolor{gray!20}4.1 & \cellcolor{gray!20}28.1 & \cellcolor{gray!20}3.6 & \cellcolor{gray!20}30.7   \\
CodeReview-New & ACR & 4.4 & 40.3 & 3.9 & 42.6   \\ \hline
\cellcolor{gray!20}\textbf{CodeReviewQA}  & \cellcolor{gray!20}ACR & \cellcolor{gray!20}4.5 & \cellcolor{gray!20}40.3 & \cellcolor{gray!20}4.1 & \cellcolor{gray!20}42.0   \\ 
~ & MCQA & \textbf{6.0} & \textbf{25.1} & \textbf{5.4} & \textbf{26.8}   \\
\hlineB{2.7}

\multicolumn{6}{l}{\textbf{ACR:} Automated Code Refinement, \textbf{NG$_5$:} 5-gram Accuracy}\\
\multicolumn{6}{l}{\textbf{MCQA}: Multiple Choice Question \& Answer, \textbf{PPL:} Perplexity Scores}\\

\end{tabular}
}

\caption{Perplexity Scores and 5-gram Accuracy (\%) of \textbf{CodeReviewQA} against existing ACR benchmarks.}
\label{table:data_contamination}
\end{table}

%% file: Tables/statistics.tex
\begin{table}[!t]
\resizebox{\columnwidth}{!}{
\begin{tabular}{llllllll}
\hlineB{2.7}
\textbf{Statistic} & \textbf{Min} & \textbf{Max} & \textbf{Mean} & \textbf{SD} & \textbf{Q1}  & \textbf{Median}  & \textbf{Q3} \\ \hline
Comment Length  &  2 &  98 &  18 &  15 &  8 &  13 &  23  \\
Code Edit Distance &  2 & 827 &  56 &  85 &  9 &  25 & 67 \\ \hline
Change Locations  & & & &  & & &  \\ 
$\phantom{0}$ Add (35)&  1 &  2 &  1 &  0 &  1 &  1 &  1 \\ 
$\phantom{0}$ Delete (144) &  1 &  19 &  3 &  4 &  1 &  2 & 4 \\ 
$\phantom{0}$ Modify (721) &  1 &  15 &  2 &  2 &  1 &  1 & 2 \\ 
\hline
Code Element Ratio  &  0.00 &  0.89 &  0.09 & 0.15 & 0.00 & 0.00 &  0.13 \\ 
Specification Ratio  &  0.04 &  165.40 &  4.88 &  9.79 &  0.67 & 1.83 &  5.07 \\\hlineB{2.7}
\multicolumn{8}{l}{See Appendix~\ref{app:benchmark_statistic} for a detailed explanation of the descriptive statistics.}
\end{tabular}
}
\caption{\textbf{CodeReviewQA} descriptive statistics.}
\label{table:statistic}
\end{table}

%% file: Figures/prompt_templates.tex
\begin{figure*}[h]
\begin{tcolorbox}[colback=gray!20, colframe=gray, 
left=0pt, right=0pt, arc=10pt, width=\textwidth, boxrule=1pt]
\fontsize{8.5}{5}\selectfont 
\textcolor{purple}{\textbf{Automated Code Refinement (ACR)}}

\vspace{1mm}
\textcolor{purple}{\{lang\} = C/CPP/CSharp/Go/Java/JavaScript/PHP/Python/Ruby}

\vspace{1mm}
\#\#\# The following \{lang\} code snippet has received a code review.

\vspace{1mm}
[\{lang\}]

\vspace{1mm}
\{code\_snippet\}

\vspace{1mm}
[/\{lang\}]

\vspace{1mm}
[CODE REVIEW]

\vspace{1mm}
\{code\_review\}

\vspace{1mm}
[/CODE REVIEW]

\vspace{1mm}
\#\#\# Please generate a revised version of the code snippet according to the code review. Do not add explanations.

\vspace{1mm}
[\{lang\}]
\vspace{1mm}

\textcolor{gray}{\textbf{ -------------------------------------------------------------------------------------------------------------------------------------------------------------- }}

\vspace{1mm}
\textcolor{purple}{\textbf{Change Type Recognition (CTR)}}

\vspace{1mm}
\textcolor{purple}{\{option\_a\}, \{option\_b\}, \{option\_c\} = only add new lines of code/only delete existing lines of code/modify the code}

\vspace{1mm}
\#\#\# The following is a multiple choice question (with answers) that tests code review comprehension. 

Question: Given this \{lang\} code snippet, what type of change is the code review asking for?

\vspace{1mm}
[\{lang\}]

\vspace{1mm}
\{code\_snippet\}

\vspace{1mm}
[/\{lang\}]

\vspace{1mm}
[CODE REVIEW]

\vspace{1mm}
\{code\_review\}

\vspace{1mm}
[/CODE REVIEW]

\vspace{1mm}
\#\#\# Possible answers:

\vspace{1mm}
A. \{option\_a\}

\vspace{1mm}
B. \{option\_b\}

\vspace{1mm}
C. \{option\_c\}

\vspace{1mm}
\#\#\# Answer with the letter symbol only. Answer:
\vspace{1mm}

\textcolor{gray}{\textbf{ -------------------------------------------------------------------------------------------------------------------------------------------------------------- }}

\vspace{1mm}
\textcolor{purple}{\textbf{Change Localisation (CL)}}

\vspace{1mm}
\textcolor{purple}{\{change\_type\} = add new lines of code under/delete code/modify code}

\vspace{1mm}
\#\#\# The following is a multiple choice question (with answers) that tests code review comprehension. 

Question: Given this \{lang\} code snippet, which line numbers is the code review asking to \{change\_type\}?

\vspace{1mm}
[\{lang\}]

\vspace{1mm}
\{code\_snippet\}

\vspace{1mm}
[/\{lang\}]

\vspace{1mm}
[CODE REVIEW]

\vspace{1mm}
\{code\_review\}

\vspace{1mm}
[/CODE REVIEW]

\vspace{1mm}
\#\#\# Possible answers:

\vspace{1mm}
A. \{option\_a\}

\vspace{1mm}
B. \{option\_b\}

\vspace{1mm}
C. \{option\_c\}

\vspace{1mm}
D. \{option\_d\}

\vspace{1mm}
\#\#\# Answer with the letter symbol only. Answer:
\vspace{1mm}

\textcolor{gray}{\textbf{ -------------------------------------------------------------------------------------------------------------------------------------------------------------- }}

\vspace{1mm}
\textcolor{purple}{\textbf{Solution Identification (SI)}}

\vspace{1mm}
\#\#\# The following is a multiple choice question (with answers) that tests code review comprehension. 

\vspace{1mm}
Question: Given this \{lang\} code snippet, which code revision is the code review asking for?

\vspace{1mm}
[\{lang\}]

\vspace{1mm}
\{code\_snippet\}

\vspace{1mm}
[/\{lang\}]

\vspace{1mm}
[CODE REVIEW]

\vspace{1mm}
\{code\_review\}

\vspace{1mm}
[/CODE REVIEW]

\vspace{1mm}
\#\#\# Possible answers:

\vspace{1mm}
A. \{option\_a\}

\vspace{1mm}
B. \{option\_b\}

\vspace{1mm}
C. \{option\_c\}

\vspace{1mm}
D. \{option\_d\}

\vspace{1mm}
\#\#\# Answer with the letter symbol only. Answer:
\vspace{1mm}

\end{tcolorbox}
\caption{Zero-Shot Prompt Templates.}
\label{fig:prompt_templates}
\end{figure*}

%% file: Figures/fewshot_CTR.tex
\begin{figure*}[h]
\begin{tcolorbox}[colback=gray!20, colframe=gray, 
left=0pt, right=0pt, arc=10pt, width=\textwidth, boxrule=1pt]
\fontsize{8.5}{5}\selectfont 
\textcolor{purple}{\textbf{Change Type Recognition (CTR)}}

\vspace{1mm}
\#\#\# The following are multiple choice questions (with answers) that tests code review comprehension. 

Question: Given this Java code snippet, what type of change is the code review asking for?

\vspace{1mm}
[Java]
\begin{lstlisting}[numbers=none, escapeinside={(*@}{@*)}, language=Java, basicstyle=\small]
   (*@\textcolor{listinggray}{1}@*)   private void launchZoomActivityAfterPermissionCheck(final View view) {
   (*@\textcolor{listinggray}{2}@*)               final Context ctx = view.getContext();
   (*@\textcolor{listinggray}{3}@*)               final Intent zoomableIntent = new Intent(ctx, ZoomableActivity.class);
   (*@\textcolor{listinggray}{4}@*)               zoomableIntent.setData(Uri.parse(media.getImageUrl()));
   (*@\textcolor{listinggray}{5}@*)               zoomableIntent.putExtra("Origin", "MediaDetail"); 
   (*@\textcolor{listinggray}{6}@*)               ctx.startActivity(
   (*@\textcolor{listinggray}{7}@*)                   zoomableIntent
   (*@\textcolor{listinggray}{8}@*)               );
\end{lstlisting}
[/Java]

\vspace{1mm}
[CODE REVIEW]

\vspace{1mm}
Could you please rename to "MediaDetails"?

\vspace{1mm}
[/CODE REVIEW]

\vspace{1mm}
\#\#\# Possible answers:

\vspace{1mm}
A. only add new lines of code

\vspace{1mm}
B. only delete existing lines of code

\vspace{1mm}
C. modify the code

\vspace{1mm}
\#\#\# Answer with the letter symbol only. Answer:

\vspace{1mm}
C

\vspace{1mm}
Question: Given this CSharp code snippet, what type of change is the code review asking for?

\vspace{1mm}
[CSharp]
\begin{lstlisting}[numbers=none, escapeinside={(*@}{@*)}, language={[Sharp]C}, basicstyle=\small]
   (*@\textcolor{listinggray}{1}@*)     public static class DateTimeDefinitions
   (*@\textcolor{listinggray}{2}@*)             public const string QuarterTypeRegex = @"(trimestral(mente)?)$";
   (*@\textcolor{listinggray}{3}@*)             public const string SemiAnnualTypeRegex = @"(semestral(mente)?)$";
   (*@\textcolor{listinggray}{4}@*)             public const string YearTypeRegex = @"(anual(mente)?)$";
   (*@\textcolor{listinggray}{5}@*)             public static readonly IList<string> FutureTerms = new List<string>
   (*@\textcolor{listinggray}{6}@*)             { 
   (*@\textcolor{listinggray}{7}@*)                 @"esea"
   (*@\textcolor{listinggray}{8}@*)             };
   (*@\textcolor{listinggray}{9}@*)         }     
   (*@\textcolor{listinggray}{10}@*)    }
   (*@\textcolor{listinggray}{11}@*)\ No newline at end of file
\end{lstlisting}
[/CSharp]

\vspace{1mm}
[CODE REVIEW]

\vspace{1mm}
There are two issues here. One, this term is not a "future" term, so we need a better name. 

Second, this value is incorrect in PT, it should be "esse", "essa", "este", "esta".

\vspace{1mm}
[/CODE REVIEW]

\vspace{1mm}
\#\#\# Possible answers:

\vspace{1mm}
A. only add new lines of code

\vspace{1mm}
B. only delete existing lines of code

\vspace{1mm}
C. modify the code

\vspace{1mm}
\#\#\# Answer with the letter symbol only. Answer:

\vspace{1mm}
C

\vspace{1mm}
Question: Given this \{lang\} code snippet, what type of change is the code review asking for?

\vspace{1mm}
[\{lang\}]

\vspace{1mm}
\{code\_snippet\}

\vspace{1mm}
[/\{lang\}]

\vspace{1mm}
[CODE REVIEW]

\vspace{1mm}
\{code\_review\}

\vspace{1mm}
[/CODE REVIEW]

\vspace{1mm}
\#\#\# Possible answers:

\vspace{1mm}
A. only add new lines of code

\vspace{1mm}
B. only delete existing lines of code

\vspace{1mm}
C. modify the code

\vspace{1mm}
\#\#\# Answer with the letter symbol only. Answer:
\vspace{1mm}

\end{tcolorbox}
\caption{Few-Shot CTR Prompt Template.}
\label{fig:fewshot_ctr}
\end{figure*}

%% file: Figures/fewshot_CL.tex
\begin{figure*}[h]
\begin{tcolorbox}[colback=gray!20, colframe=gray, 
left=0pt, right=0pt, arc=10pt, width=\textwidth, boxrule=1pt]
\fontsize{8.5}{5}\selectfont 
\textcolor{purple}{\textbf{Change Localisation (CL)}}

\vspace{1mm}
\textcolor{purple}{\{change\_type\} = add new lines of code under/delete code/modify code}

\vspace{1mm}
\#\#\# The following are multiple choice questions (with answers) that tests code review comprehension. 

Question: Given this Java code snippet, which line numbers is the code review asking to modify code?

\vspace{1mm}
[Java]
\begin{lstlisting}[numbers=none, escapeinside={(*@}{@*)}, language=Java, basicstyle=\small]
   (*@\textcolor{listinggray}{1}@*)   private void launchZoomActivityAfterPermissionCheck(final View view) {
   (*@\textcolor{listinggray}{2}@*)               final Context ctx = view.getContext();
   (*@\textcolor{listinggray}{3}@*)               final Intent zoomableIntent = new Intent(ctx, ZoomableActivity.class);
   (*@\textcolor{listinggray}{4}@*)               zoomableIntent.setData(Uri.parse(media.getImageUrl()));
   (*@\textcolor{listinggray}{5}@*)               zoomableIntent.putExtra("Origin", "MediaDetail"); 
   (*@\textcolor{listinggray}{6}@*)               ctx.startActivity(
   (*@\textcolor{listinggray}{7}@*)                   zoomableIntent
   (*@\textcolor{listinggray}{8}@*)               );
\end{lstlisting}
[/Java]

\vspace{1mm}
[CODE REVIEW]

\vspace{1mm}
Could you please rename to "MediaDetails"?

\vspace{1mm}
[/CODE REVIEW]

\vspace{1mm}
\#\#\# Possible answers:

\vspace{1mm}
A. 3

\vspace{1mm}
B. 4

\vspace{1mm}
C. 5

\vspace{1mm}
D. 6

\vspace{1mm}
\#\#\# Answer with the letter symbol only. Answer:

\vspace{1mm}
C

\vspace{1mm}
Question: Given this CSharp code snippet, which line numbers is the code review asking to modify code?

\vspace{1mm}
[CSharp]
\begin{lstlisting}[numbers=none, escapeinside={(*@}{@*)}, language={[Sharp]C}, basicstyle=\small]
   (*@\textcolor{listinggray}{1}@*)     public static class DateTimeDefinitions
   (*@\textcolor{listinggray}{2}@*)             public const string QuarterTypeRegex = @"(trimestral(mente)?)$";
   (*@\textcolor{listinggray}{3}@*)             public const string SemiAnnualTypeRegex = @"(semestral(mente)?)$";
   (*@\textcolor{listinggray}{4}@*)             public const string YearTypeRegex = @"(anual(mente)?)$";
   (*@\textcolor{listinggray}{5}@*)             public static readonly IList<string> FutureTerms = new List<string>
   (*@\textcolor{listinggray}{6}@*)             { 
   (*@\textcolor{listinggray}{7}@*)                 @"esea"
   (*@\textcolor{listinggray}{8}@*)             };
   (*@\textcolor{listinggray}{9}@*)         }     
   (*@\textcolor{listinggray}{10}@*)    }
   (*@\textcolor{listinggray}{11}@*)\ No newline at end of file
\end{lstlisting}
[/CSharp]

\vspace{1mm}
[CODE REVIEW]

\vspace{1mm}
There are two issues here. One, this term is not a "future" term, so we need a better name. 

Second, this value is incorrect in PT, it should be "esse", "essa", "este", "esta".

\vspace{1mm}
[/CODE REVIEW]

\vspace{1mm}
\#\#\# Possible answers:

\vspace{1mm}
A. 5, 7

\vspace{1mm}
B. 2, 3

\vspace{1mm}
C. 1, 4

\vspace{1mm}
D. 8, 11

\vspace{1mm}
\#\#\# Answer with the letter symbol only. Answer:

\vspace{1mm}
A

\vspace{1mm}
Question: Given this \{lang\} code snippet, which line numbers is the code review asking to \{change\_type\}?

\vspace{1mm}
[\{lang\}]

\vspace{1mm}
\{code\_snippet\}

\vspace{1mm}
[/\{lang\}]

\vspace{1mm}
[CODE REVIEW]

\vspace{1mm}
\{code\_review\}

\vspace{1mm}
[/CODE REVIEW]

\vspace{1mm}
\#\#\# Possible answers:

\vspace{1mm}
A. \{option\_a\}

\vspace{1mm}
B. \{option\_b\}

\vspace{1mm}
C. \{option\_c\}

\vspace{1mm}
D. \{option\_d\}

\vspace{1mm}
\#\#\# Answer with the letter symbol only. Answer:
\vspace{1mm}

\end{tcolorbox}
\caption{Few-Shot CL Prompt Template.}
\label{fig:fewshot_cl}
\end{figure*}

%% file: Figures/fewshot_SI.tex
\begin{figure*}[h]
\begin{tcolorbox}[colback=gray!20, colframe=gray, 
left=0pt, right=0pt, arc=10pt, width=\textwidth, boxrule=1pt]
\fontsize{7.5}{5}\selectfont 
\textcolor{purple}{\textbf{Solution Identification (SI)}}

\vspace{1mm}
\#\#\# The following are multiple choice questions (with answers) that tests code review comprehension. 

Question: Given this Java code snippet, which code revision is the code review asking for?

\vspace{1mm}
[Java]
\begin{lstlisting}[numbers=none, escapeinside={(*@}{@*)}, language=Java, basicstyle=\scriptsize]
   (*@\textcolor{listinggray}{1}@*)   private void launchZoomActivityAfterPermissionCheck(final View view) {
   (*@\textcolor{listinggray}{2}@*)               final Context ctx = view.getContext();
   (*@\textcolor{listinggray}{3}@*)               final Intent zoomableIntent = new Intent(ctx, ZoomableActivity.class);
   (*@\textcolor{listinggray}{4}@*)               zoomableIntent.setData(Uri.parse(media.getImageUrl()));
   (*@\textcolor{listinggray}{5}@*)               zoomableIntent.putExtra("Origin", "MediaDetail"); 
   (*@\textcolor{listinggray}{6}@*)               ctx.startActivity(
   (*@\textcolor{listinggray}{7}@*)                   zoomableIntent
   (*@\textcolor{listinggray}{8}@*)               );
\end{lstlisting}
[/Java]

\vspace{1mm}
[CODE REVIEW]

\vspace{1mm}
Could you please rename to "MediaDetails"?

\vspace{1mm}
[/CODE REVIEW]

\vspace{1mm}
\#\#\# Possible answers:

\vspace{1mm}
A.
\begin{lstlisting}[numbers=none, escapeinside={(*@}{@*)}, language=Java, basicstyle=\scriptsize]
   (*@\textcolor{listinggray}{5}@*)- zoomableIntent.putExtra("Origin", "MediaDetail"); {   (*@\textcolor{listinggray}{5}@*)+ zoomableIntent.postExtra("Origin", "MediaDetails");
\end{lstlisting}

\vspace{1mm}
B.
\begin{lstlisting}[numbers=none, escapeinside={(*@}{@*)}, language=Java, basicstyle=\scriptsize]
   (*@\textcolor{listinggray}{5}@*)- zoomableIntent.putExtra("Origin", "MediaDetail"); {   (*@\textcolor{listinggray}{5}@*)+ zoomableIntent.getExtra("Origin", "MediaDetails");
\end{lstlisting}

\vspace{1mm}
C.
\begin{lstlisting}[numbers=none, escapeinside={(*@}{@*)}, language=Java, basicstyle=\scriptsize]
   (*@\textcolor{listinggray}{5}@*)- zoomableIntent.putExtra("Origin", "MediaDetail"); {   (*@\textcolor{listinggray}{5}@*)+ zoomableIntent.putExtra("Origin", "MediaDetails");
\end{lstlisting}

\vspace{1mm}
D.
\begin{lstlisting}[numbers=none, escapeinside={(*@}{@*)}, language=Java, basicstyle=\scriptsize]
   (*@\textcolor{listinggray}{5}@*)- zoomableIntent.putExtra("Origin", "MediaDetail"); {   (*@\textcolor{listinggray}{5}@*)+ zoomableIntent.renameExtra("Origin", "MediaDetails");
\end{lstlisting}

\vspace{1mm}
\#\#\# Answer with the letter symbol only. Answer:

\vspace{1mm}
C

\vspace{1mm}
Question: Given this CSharp code snippet, which code revision is the code review asking for?

\vspace{1mm}
[CSharp]
\begin{lstlisting}[numbers=none, escapeinside={(*@}{@*)}, language={[Sharp]C}, basicstyle=\scriptsize]
   (*@\textcolor{listinggray}{1}@*)     public static class DateTimeDefinitions
   (*@\textcolor{listinggray}{2}@*)             public const string QuarterTypeRegex = @"(trimestral(mente)?)$";
   (*@\textcolor{listinggray}{3}@*)             public const string SemiAnnualTypeRegex = @"(semestral(mente)?)$";
   (*@\textcolor{listinggray}{4}@*)             public const string YearTypeRegex = @"(anual(mente)?)$";
   (*@\textcolor{listinggray}{5}@*)             public static readonly IList<string> FutureTerms = new List<string>
   (*@\textcolor{listinggray}{6}@*)             { 
   (*@\textcolor{listinggray}{7}@*)                 @"esea"
   (*@\textcolor{listinggray}{8}@*)             };
   (*@\textcolor{listinggray}{9}@*)         }     
   (*@\textcolor{listinggray}{10}@*)    }
   (*@\textcolor{listinggray}{11}@*)\ No newline at end of file
\end{lstlisting}
[/CSharp]

\vspace{1mm}
[CODE REVIEW]

\vspace{1mm}
There are two issues here. One, this term is not a "future" term, so we need a better name. 

Second, this value is incorrect in PT, it should be "esse", "essa", "este", "esta".

\vspace{1mm}
[/CODE REVIEW]

\vspace{1mm}
\#\#\# Possible answers:

\vspace{1mm}
A.
\begin{lstlisting}[numbers=none, escapeinside={(*@}{@*)}, language=Java, basicstyle=\scriptsize]
   (*@\textcolor{listinggray}{5}@*)- public static readonly IList<string> FutureTerms = new List<string>
   (*@\textcolor{listinggray}{5}@*)+ public static readonly IList<string> ThisTerms = new List<string>
   (*@\textcolor{listinggray}{7}@*)- @"esea" (*@\textcolor{listinggray}{7}@*)+ @"esse" (*@\textcolor{listinggray}{8}@*)+ @"essa" (*@\textcolor{listinggray}{9}@*)+ @"este" (*@\textcolor{listinggray}{10}@*)+ @"esta"
\end{lstlisting}

\vspace{1mm}
B.
\begin{lstlisting}[numbers=none, escapeinside={(*@}{@*)}, language=Java, basicstyle=\scriptsize]
   (*@\textcolor{listinggray}{5}@*)- public static readonly IList<string> FutureTerms = new List<string>
   (*@\textcolor{listinggray}{5}@*)+ public static readonly IList<string> NotTerms = new List<string>
   (*@\textcolor{listinggray}{7}@*)- @"esea" (*@\textcolor{listinggray}{7}@*)+ @"esse" (*@\textcolor{listinggray}{8}@*)+ @"essa" (*@\textcolor{listinggray}{9}@*)+ @"este" (*@\textcolor{listinggray}{10}@*)+ @"esta"
\end{lstlisting}

\vspace{1mm}
C.
\begin{lstlisting}[numbers=none, escapeinside={(*@}{@*)}, language=Java, basicstyle=\scriptsize]
   (*@\textcolor{listinggray}{5}@*)- public static readonly IList<string> FutureTerms = new List<string>
   (*@\textcolor{listinggray}{5}@*)+ public static readonly IList<string> futureTerms = new List<string>
   (*@\textcolor{listinggray}{7}@*)- @"esea" (*@\textcolor{listinggray}{7}@*)+ @"esse" (*@\textcolor{listinggray}{8}@*)+ @"essa" (*@\textcolor{listinggray}{9}@*)+ @"este" (*@\textcolor{listinggray}{10}@*)+ @"esta"
\end{lstlisting}

\vspace{1mm}
D.
\begin{lstlisting}[numbers=none, escapeinside={(*@}{@*)}, language=Java, basicstyle=\scriptsize]
   (*@\textcolor{listinggray}{5}@*)- public static readonly IList<string> FutureTerms = new List<string>
   (*@\textcolor{listinggray}{5}@*)+ public static readonly IList<string> betterTerms = new List<string>
   (*@\textcolor{listinggray}{7}@*)- @"esea" (*@\textcolor{listinggray}{7}@*)+ @"esse" (*@\textcolor{listinggray}{8}@*)+ @"essa" (*@\textcolor{listinggray}{9}@*)+ @"este" (*@\textcolor{listinggray}{10}@*)+ @"esta"
\end{lstlisting}

\vspace{1mm}
\#\#\# Answer with the letter symbol only. Answer:

\vspace{1mm}
A

\vspace{1mm}
Question: Given this \{lang\} code snippet, which code revision is the code review asking for?

\vspace{1mm}
[\{lang\}]

\vspace{1mm}
\{code\_snippet\}

\vspace{1mm}
[/\{lang\}]

\vspace{1mm}
[CODE REVIEW]

\vspace{1mm}
\{code\_review\}

\vspace{1mm}
[/CODE REVIEW]

\vspace{1mm}
\#\#\# Possible answers:

\vspace{1mm}
A. \{option\_a\}

\vspace{1mm}
B. \{option\_b\}

\vspace{1mm}
C. \{option\_c\}

\vspace{1mm}
D. \{option\_d\}

\vspace{1mm}
\#\#\# Answer with the letter symbol only. Answer:
\vspace{1mm}

\end{tcolorbox}
\caption{Few-Shot SI Prompt Template.}
\label{fig:fewshot_si}
\end{figure*}

%% file: Figures/difficulty_variation.tex
\begin{figure*}[h]
\begin{tcolorbox}[colback=gray!20, colframe=gray, 
left=0pt, right=0pt, arc=10pt, width=\linewidth, boxrule=1pt]
\fontsize{9}{5}\selectfont 
\textcolor{purple}{\textbf{Pre-Review Code Submission}}

\vspace{1mm}
\begin{lstlisting}[numbers=none, escapeinside={(*@}{@*)}, language=Python, basicstyle=\scriptsize]
   (*@\textcolor{listinggray}{1}@*)       def main(args: argparse.Namespace):
   (*@\textcolor{listinggray}{2}@*)                 )
   (*@\textcolor{listinggray}{3}@*)             host_environment = host_environments.pop()
   (*@\textcolor{listinggray}{4}@*)
   (*@\textcolor{listinggray}{5}@*)             module_dir_paths = sort_and_dedup_paths([
   (*@\textcolor{listinggray}{6}@*)                 iree_artifacts.get_module_dir_path(config.module_generation_config)
   (*@\textcolor{listinggray}{7}@*)                 for config in run_configs
   (*@\textcolor{listinggray}{8}@*)             ])
   (*@\textcolor{listinggray}{9}@*)
   (*@\textcolor{listinggray}{10}@*)            output_map[device_name] = {
   (*@\textcolor{listinggray}{11}@*)                "host_environment": dataclasses.asdict(host_environment),
\end{lstlisting}
\vspace{2mm}

\textcolor{purple}{\textbf{Code Review:}} 
Huh, would be nice if the path was just naturally serializable
\vspace{2mm}

\textcolor{gray}{\textbf{ ------------------------------------------------------------------------------------------------------------------------------------------------------ }}

\vspace{1mm}
\textcolor{purple}{\textbf{Which line numbers is the code review asking to modify code?}} 
\textcolor{dkgreen}{A. $\checkmark$} line numbers 5, 6, 8 

\vspace{1mm}
\textcolor{purple}{Change Localisation (Easy)}

\vspace{1mm}
B. line numbers 1, 2, 3$\phantom{\checkmark}$ C. line numbers 4, 7, 9$\phantom{\checkmark}$  D. line numbers 9, 10, 11$\phantom{\checkmark}$ 

\vspace{1mm}
\textcolor{purple}{Change Localisation (Hard)}

\vspace{1mm}
B. line numbers 1, 5, 6$\phantom{\checkmark}$ C. line numbers 3, 5, 6$\phantom{\checkmark}$  D. line numbers 3, 5, 8$\phantom{\checkmark}$
\vspace{2mm}

\textcolor{gray}{\textbf{ ------------------------------------------------------------------------------------------------------------------------------------------------------ }}

\vspace{1mm}
\textcolor{purple}{\textbf{Which code revision is the code review asking for?}} 

\vspace{1mm}
\textcolor{dkgreen}{A. $\checkmark$}
\begin{lstlisting}[numbers=none, escapeinside={(*@}{@*)}, language=Python, language=C, basicstyle=\scriptsize]
   (*@\textcolor{listinggray}{5}@*)  -    module_dir_paths = sort_and_dedup_paths([
   (*@\textcolor{listinggray}{5}@*)  +    module_dir_paths = sorted(set(
   (*@\textcolor{listinggray}{6}@*)  -        iree_artifacts.get_module_dir_path(config.module_generation_config)
   (*@\textcolor{listinggray}{6}@*)  +        str(iree_artifacts.get_module_dir_path(config.module_generation_config))
   (*@\textcolor{listinggray}{8}@*)  -    ])
   (*@\textcolor{listinggray}{8}@*)  +    ))
\end{lstlisting}

\vspace{1mm}
\textcolor{purple}{Solution Identification (Easy)}

\vspace{1mm}
B. 
\begin{lstlisting}[numbers=none, escapeinside={(*@}{@*)}, language=Python, language=C,basicstyle=\scriptsize]
   (*@\textcolor{listinggray}{5}@*)  -    module_dir_paths = sort_and_dedup_paths([
   (*@\textcolor{listinggray}{5}@*)  +    module_dir_paths = sorted(set(
   (*@\textcolor{listinggray}{6}@*)  -        iree_artifacts.get_module_dir_path(config.module_generation_config)
   (*@\textcolor{listinggray}{6}@*)  +        str(struct_lucule.get_module_dir_path(config.module_generation_config))
   (*@\textcolor{listinggray}{8}@*)  -    ])
   (*@\textcolor{listinggray}{8}@*)  +    ))
\end{lstlisting}

\vspace{1mm}
C. 
\begin{lstlisting}[numbers=none, escapeinside={(*@}{@*)}, language=Python, language=C,basicstyle=\scriptsize]
   (*@\textcolor{listinggray}{5}@*)  -    module_dir_paths = sort_and_dedup_paths([
   (*@\textcolor{listinggray}{5}@*)  +    module_dir_paths = sorted(set(
   (*@\textcolor{listinggray}{6}@*)  -        iree_artifacts.get_module_dir_path(config.module_generation_config)
   (*@\textcolor{listinggray}{6}@*)  +        str(assertTrue_localhost.get_module_dir_path(config.module_generation_config))
   (*@\textcolor{listinggray}{8}@*)  -    ])
   (*@\textcolor{listinggray}{8}@*)  +    ))
\end{lstlisting}

\vspace{1mm}
D. 
\begin{lstlisting}[numbers=none, escapeinside={(*@}{@*)}, language=Python, language=C,basicstyle=\scriptsize]
   (*@\textcolor{listinggray}{5}@*)  -    module_dir_paths = sort_and_dedup_paths([
   (*@\textcolor{listinggray}{5}@*)  +    module_dir_paths = sorted(set(
   (*@\textcolor{listinggray}{6}@*)  -        iree_artifacts.get_module_dir_path(config.module_generation_config)
   (*@\textcolor{listinggray}{6}@*)  +        str(indexChat_retry.get_module_dir_path(config.module_generation_config))
   (*@\textcolor{listinggray}{8}@*)  -    ])
   (*@\textcolor{listinggray}{8}@*)  +    ))
\end{lstlisting}

\vspace{1mm}
\textcolor{purple}{Solution Identification (Hard)}

\vspace{1mm}
B. 
\begin{lstlisting}[numbers=none, escapeinside={(*@}{@*)}, language=Python, language=C,basicstyle=\scriptsize]
   (*@\textcolor{listinggray}{5}@*)  -    module_dir_paths = sort_and_dedup_paths([
   (*@\textcolor{listinggray}{5}@*)  +    module_dir_paths = sorted(set(
   (*@\textcolor{listinggray}{6}@*)  -        iree_artifacts.get_module_dir_path(config.module_generation_config)
   (*@\textcolor{listinggray}{6}@*)  +        str(View_DEF.get_module_dir_path(config.module_generation_config))
   (*@\textcolor{listinggray}{8}@*)  -    ])
   (*@\textcolor{listinggray}{8}@*)  +    ))
\end{lstlisting}

\vspace{1mm}
C. 
\begin{lstlisting}[numbers=none, escapeinside={(*@}{@*)}, language=Python, language=C,basicstyle=\scriptsize]
   (*@\textcolor{listinggray}{5}@*)  -    module_dir_paths = sort_and_dedup_paths([
   (*@\textcolor{listinggray}{5}@*)  +    module_dir_paths = sorted(set(
   (*@\textcolor{listinggray}{6}@*)  -        iree_artifacts.get_module_dir_path(config.module_generation_config)
   (*@\textcolor{listinggray}{6}@*)  +        str(develop_weight.get_module_dir_path(config.module_generation_config))
   (*@\textcolor{listinggray}{8}@*)  -    ])
   (*@\textcolor{listinggray}{8}@*)  +    ))
\end{lstlisting}

\vspace{1mm}
D. 
\begin{lstlisting}[numbers=none, escapeinside={(*@}{@*)}, language=Python, language=C,basicstyle=\scriptsize]
   (*@\textcolor{listinggray}{5}@*)  -    module_dir_paths = sort_and_dedup_paths([
   (*@\textcolor{listinggray}{5}@*)  +    module_dir_paths = sorted(set(
   (*@\textcolor{listinggray}{6}@*)  -        iree_artifacts.get_module_dir_path(config.module_generation_config)
   (*@\textcolor{listinggray}{6}@*)  +        str(register_access.get_module_dir_path(config.module_generation_config))
   (*@\textcolor{listinggray}{8}@*)  -    ])
   (*@\textcolor{listinggray}{8}@*)  +    ))
\end{lstlisting}

\end{tcolorbox}
\caption{Examples of Variation in Difficulty.}
\label{fig:difficulty_variation}
\end{figure*}

%% file: Figures/comment_length.tex
\begin{figure*}[h]
\begin{tcolorbox}[colback=gray!20, colframe=gray, 
left=0pt, right=0pt, arc=10pt, width=\textwidth, boxrule=1pt]
\fontsize{9}{5}\selectfont 
\textcolor{purple}{\textbf{Comment Length = 8}}

FYI, this will spam console when running `aaa`.
\vspace{2mm}

\textcolor{gray}{\textbf{ ------------------------------------------------------------------------------------------------------------------------------------------------------ }}

\vspace{1mm}
\textcolor{purple}{\textbf{Comment Length = 18}}

By the format string it looks like parameters shall be reversed. Type shall be 1st and exception 2nd
\vspace{2mm}

\textcolor{gray}{\textbf{ ------------------------------------------------------------------------------------------------------------------------------------------------------ }}

\vspace{1mm}
\textcolor{purple}{\textbf{Comment Length = 23}}

I'm wondering if it's useful to show the message from the exception in this debug message, at least in the case of IOException.

\end{tcolorbox}
\caption{Examples of Different Comment Lengths.}
\label{fig:comment_length}
\end{figure*}

%% file: Figures/code_edit_distance.tex
\begin{figure*}[h]
\begin{tcolorbox}[colback=gray!20, colframe=gray, 
left=0pt, right=0pt, arc=10pt, width=\linewidth, boxrule=1pt]
\fontsize{9}{5}\selectfont 
\textcolor{purple}{\textbf{Code Edit Distance = 9}}

\vspace{1mm}
\begin{lstlisting}[numbers=none, escapeinside={(*@}{@*)}, language=C,basicstyle=\small]
   (*@\textcolor{listinggray}{1}@*)     OnConflictAction onconflict_action = ts_chunk_dispatch_get_on_conflict_action(dispatch);
   (*@\textcolor{listinggray}{2}@*)     ResultRelInfo *resrelinfo, *relinfo;
   (*@\textcolor{listinggray}{3}@*)  -	bool has_compressed_chunk = (chunk->fd.compressed_chunk_id != 0);
   (*@\textcolor{listinggray}{3}@*)  +	bool is_compressed = (chunk->fd.compressed_chunk_id != 0)
   (*@\textcolor{listinggray}{4}@*)     /* permissions NOT checked here; were checked at hypertable level */
   (*@\textcolor{listinggray}{5}@*)     if (check_enable_rls(chunk->table_id, InvalidOid, false) == RLS_ENABLED)
\end{lstlisting}
\vspace{2mm}

\textcolor{purple}{\textbf{Code Review:}} 
Should `is\_compressed` now be given by the chunk's compression status flag? Otherwise it looks like we have different ways of determining compression status.
\vspace{2mm}

\textcolor{gray}{\textbf{ ------------------------------------------------------------------------------------------------------------------------------------------------------ }}

\vspace{1mm}
\textcolor{purple}{\textbf{Code Edit Distance = 25}}
\vspace{1mm}
\begin{lstlisting}[numbers=none, escapeinside={(*@}{@*)}, language=C++, language=C,basicstyle=\small]
   (*@\textcolor{listinggray}{1}@*)     void Server_Card::resetState()
   (*@\textcolor{listinggray}{2}@*)     setPT(QString());
   (*@\textcolor{listinggray}{3}@*)     setAnnotation(QString());
   (*@\textcolor{listinggray}{4}@*)     setDoesntUntap(false);
   (*@\textcolor{listinggray}{5}@*)  -    setFaceDown(false);
   (*@\textcolor{listinggray}{6}@*)     }
   (*@\textcolor{listinggray}{7}@*)      
   (*@\textcolor{listinggray}{8}@*)     QString Server_Card::setAttribute(CardAttribute attribute, const QString &avalue, bool allCards)
\end{lstlisting}
\vspace{2mm}

\textcolor{purple}{\textbf{Code Review:}} 
this causes a major bug: cards have their state reset when moved between the battlefield and the deck, their facedown state is then checked afterwards to determine what event to show to other players. with this change moving a facedown card to your deck (unknown to unknown) will tell all your opponents (but not you) what card it was.
\vspace{2mm}

\textcolor{gray}{\textbf{ ------------------------------------------------------------------------------------------------------------------------------------------------------ }}

\vspace{1mm}
\textcolor{purple}{\textbf{Code Edit Distance = 67}}
\vspace{1mm}
\begin{lstlisting}[numbers=none, escapeinside={(*@}{@*)}, language={[Sharp]C}, language=C,basicstyle=\small]
   (*@\textcolor{listinggray}{1}@*)     public unsafe LazyStringValue GetDocumentId(LazyStringValue key)
   (*@\textcolor{listinggray}{2}@*)                 if (index == -1)
   (*@\textcolor{listinggray}{3}@*)                     return null;
   (*@\textcolor{listinggray}{4}@*)
   (*@\textcolor{listinggray}{5}@*)   -             _tmpLazyStringInstance = _context.GetLazyString(key.Buffer, index);
   (*@\textcolor{listinggray}{5}@*)   +             return _context.GetLazyString(key.Buffer, index); 
   (*@\textcolor{listinggray}{6}@*)   -             return _tmpLazyStringInstance;
   (*@\textcolor{listinggray}{7}@*)         }
   (*@\textcolor{listinggray}{8}@*)
   (*@\textcolor{listinggray}{9}@*)         // TODO unify if possible with AllowedPathsValidator
\end{lstlisting}
\vspace{2mm}

\textcolor{purple}{\textbf{Code Review:}} 
Why do you store that in the temporary variable?

\end{tcolorbox}
\caption{Examples of Different Code Edit Distances.}
\label{fig:code_edit_distance}
\end{figure*}

%% file: Figures/code_element_ratio.tex
\begin{figure*}[h]
\begin{tcolorbox}[colback=gray!20, colframe=gray, 
left=0pt, right=0pt, arc=10pt, width=\textwidth, boxrule=1pt]
\fontsize{9}{5}\selectfont 
\textcolor{purple}{\textbf{Code Element Ratio = 0}}

\vspace{1mm}
Should this really be a compile time error? The fact that is can be imported multiple times does not mean that it will be.
\vspace{2mm}

\textcolor{gray}{\textbf{ ------------------------------------------------------------------------------------------------------------------------------------------------------ }}

\vspace{1mm}
\textcolor{purple}{\textbf{Code Element Ratio = 0.13}}

\vspace{1mm}
For all of the fuzz tests, does it make sense to have versions for `len\_prefixed` both `true` and `false` ?
\vspace{2mm}

\textcolor{gray}{\textbf{ ------------------------------------------------------------------------------------------------------------------------------------------------------ }}

\vspace{1mm}
\textcolor{purple}{\textbf{Code Element Ratio = 0.37}}

\vspace{1mm}
I think you're missing a `flb\_free(seq\_index\_str);` there.

Other than that, would you mind change that comparison to `if (tmp\_key == NULL) \{` instead? I'd really appreciate it.

\end{tcolorbox}
\caption{Examples of Different Code Element Ratios.}
\label{fig:code_element_ratio}
\end{figure*}

%% file: Figures/spec_ratio.tex
\begin{figure*}[h]
\begin{tcolorbox}[colback=gray!20, colframe=gray, 
left=0pt, right=0pt, arc=10pt, width=\linewidth, boxrule=1pt]
\fontsize{9}{5}\selectfont 
\textcolor{purple}{\textbf{Specification Ratio = 0.67}}

\vspace{1mm}
\begin{lstlisting}[numbers=none, escapeinside={(*@}{@*)}, language=C++,basicstyle=\small]
   (*@\textcolor{listinggray}{1}@*)     void hpx_thread_buffer::resize(const std::size_t num_threads,
   (*@\textcolor{listinggray}{2}@*)     }
   (*@\textcolor{listinggray}{3}@*)
   (*@\textcolor{listinggray}{4}@*)     void *hpx_thread_buffer::get(std::size_t thread_num) const noexcept {
   (*@\textcolor{listinggray}{5}@*)  -     KOKKOS_ASSERT(thread_num < m_num_threads);
   (*@\textcolor{listinggray}{5}@*)  +     KOKKOS_EXPECTS(thread_num < m_num_threads);
   (*@\textcolor{listinggray}{6}@*)        if (m_data == nullptr) {
   (*@\textcolor{listinggray}{7}@*)             return nullptr;
   (*@\textcolor{listinggray}{8}@*)        }
   (*@\textcolor{listinggray}{9}@*)        return &m_data[thread_num * m_size_per_thread];
   (*@\textcolor{listinggray}{10}@*)     }
   (*@\textcolor{listinggray}{11}@*)
   (*@\textcolor{listinggray}{12}@*)    void *hpx_thread_buffer::get_extra_space() const noexcept {
   (*@\textcolor{listinggray}{13}@*) -     KOKKOS_ASSERT(m_extra_space > 0);
   (*@\textcolor{listinggray}{13}@*) +     KOKKOS_EXPECTS(m_extra_space > 0);
   (*@\textcolor{listinggray}{14}@*)       if (m_data == nullptr) {
   (*@\textcolor{listinggray}{15}@*)            return nullptr;
   (*@\textcolor{listinggray}{16}@*)       }
\end{lstlisting}
\vspace{1mm}

\textcolor{purple}{\textbf{Code Review:}} 
This is fine but just pointing out there is also a `KOKKOS\_EXPECTS` that was meant for checking preconditions
\vspace{2mm}

\textcolor{gray}{\textbf{ ------------------------------------------------------------------------------------------------------------------------------------------------------ }}

\vspace{1mm}
\textcolor{purple}{\textbf{Specification Ratio = 37.60}}
\vspace{1mm}
\begin{lstlisting}[numbers=none, escapeinside={(*@}{@*)}, language=Java,basicstyle=\small]
   (*@\textcolor{listinggray}{1}@*)     private String addNashornJavaScriptEngineIfNecessary(String cp) {
   (*@\textcolor{listinggray}{2}@*)          }
   (*@\textcolor{listinggray}{3}@*)
   (*@\textcolor{listinggray}{4}@*)         private boolean requiresNashornJavaScriptEngine() {
   (*@\textcolor{listinggray}{5}@*)  -          String version = System.getProperty("java.specification.version");
   (*@\textcolor{listinggray}{5}@*)  +          return getJavaVersion() >= 15; // Nashorn was removed in Java 15
   (*@\textcolor{listinggray}{6}@*)  -          if (version.startsWith("1.")) {
   (*@\textcolor{listinggray}{7}@*)  -              version = version.substring(2);      
   (*@\textcolor{listinggray}{8}@*)  -          }
   (*@\textcolor{listinggray}{9}@*)  -          return Integer.parseInt(version) >= 15; // Nashorn was removed in Java 15
   (*@\textcolor{listinggray}{10}@*)         }
   (*@\textcolor{listinggray}{11}@*)
   (*@\textcolor{listinggray}{12}@*)     }      
\end{lstlisting}
\vspace{1mm}

\textcolor{purple}{\textbf{Code Review:}} 
You can use `getJavaVersion()` here.

\end{tcolorbox}
\caption{Examples of Different Specification Ratios.}
\label{fig:specification_ratio}
\end{figure*}

%% file: Tables/advanced_prompting.tex
\begin{table}[!t]

\resizebox{\columnwidth}{!}{
\begin{tabular}{lccccc}
\hlineB{2.7}
\textbf{Prompt Strategy} & \textbf{CTR} & \textbf{CL$_\text{E}$} & \textbf{CL$_\text{H}$} & \textbf{SI$_\text{E}$} & \textbf{SI$_\text{H}$} \\ \hline

\cellcolor{gray!20}Zero-Shot &\cellcolor{gray!20}68.4 &\cellcolor{gray!20}74.7 &\cellcolor{gray!20}\textbf{69.0} &\cellcolor{gray!20}\textbf{84.2} &\cellcolor{gray!20}\textbf{76.7} \\

\hline

One-Shot	&74.1	&68.6	&61.9	&59.9	&55.6 \\ 

\cellcolor{gray!20}Two-Shot &\cellcolor{gray!20}\textbf{76.9} &  \cellcolor{gray!20}\textbf{75.8} & \cellcolor{gray!20}68.7 & \cellcolor{gray!20}57.3 & \cellcolor{gray!20}54.6 \\

Chain-of-Thought &68.6	&65.8	&60.1	&66.6	&60.0 \\ 

\hlineB{2.7}

\multicolumn{6}{l}{\textbf{CTR:} Change Type Recognition, \textbf{E:} Easy, \textbf{H:} Hard} \\
\multicolumn{6}{l}{\textbf{CL:} Change Localisation, \textbf{SI:} Solution Identification} \\
\end{tabular}
}

\caption{MCQA Probe Accuracy (\%) of Llama-3.1-70B-Instruct w/ Advanced Prompting.}
\label{table:advanced_prompting}\
\end{table}

%% file: Figures/si_distract_alg.tex
\begin{algorithm*}[h]
\caption{Create $H_{post^-}$ Distractors for Solution Identification}
\label{alg:mutant}
\SetAlgoLined
\KwIn{Surrogate LLM $\textcolor{purple}{f_\theta}$, Temperature $\textcolor{teal}{k}$, No. of Distractors $\textcolor{orange}{N}$, Difficulty $\textcolor{brown}{\phi}$}
\KwOut{Set of $H_{post^-}$ Distractors $\textcolor{violet}{D}$}

\textcolor{blue}{\tcp{Identify code elements in the changed lines of the post-review code revision}}

$Lines \gets \text{GetChangedLines}(H_{post^+})$\;  
$AST \gets \text{GetAbstractSyntaxTree}(H_{post^+})$\; 
$Nodes\gets \text{GetLeafNodes}(AST, Lines)$\;

\textcolor{blue}{\tcp{Calculate average token surprisal for each identified code element}}
$S_{token}, S_{node}, C_{distractor}, \textcolor{violet}{D} \gets \emptyset, \emptyset, \emptyset, \emptyset$\;

\For{$c_t \in H_{post^+}$}{ 
    $\bm{h}_{t-1} \gets \textcolor{purple}{f_\theta(}H_{pre},R_{nl},c_{<t}\textcolor{purple}{)}$\;
    $S_{token}[c_t] \gets -\log_2 P(c_t|\bm{h}_{t-1})$\;
    }

\For{$n_t \in Nodes$}{ 
    \For{$c_t \in n_t$}{ 
        $S_{node}[n_t] \gets S_{node}[n_t] + S_{token}[c_t]$\;
    }
    $S_{node}[n_t] \gets \frac{S_{node}[n_t]}{length(n_t)}$\;
}

\textcolor{blue}{\tcp{Mask the code element with the highest average token surprisal}}
$n_{max} \gets \text{GetMaxKeys}(S_{node}, 1)$\;
$Mask \gets \text{ApplyMask}(H_{post^+}, n_{max})$

\textcolor{blue}{\tcp{Create a set of distractors for each of the easy and hard variations}}
\While{$length(C_{distractor})<2 \times \textcolor{orange}{N}$}{
    $\hat{n}_{max} \gets \arg\max\limits_{n_{max}} P_{\textcolor{teal}{k}}(n_{max}|\textcolor{purple}{f_\theta(}Mask\textcolor{purple}{)})$\;
    $Candidate = \text{InFill}(Mask, \hat{n}_{max})$\;
    
    \If{$Candidate \neq H_{post^+} \land Candidate \notin C_{distractor}$}{
        $\theta = \text{Cosine}(\textcolor{purple}{f_\theta(}Candidate\textcolor{purple}{)},\textcolor{purple}{f_\theta(}H_{post^+}\textcolor{purple}{)})$\;
        $C_{distractor}[Candidate] \gets \theta$\;
    }
}

\textcolor{blue}{\tcp{Select distractors most semantically different from ground truth for easy and distractors most semantically similar to ground truth for hard}}
\If{$\textcolor{brown}{\phi} = easy$}{
    $\textcolor{violet}{D} \gets \text{GetMinKeys}(C_{distractor}, \textcolor{orange}{N})$\;
}
\ElseIf{$\textcolor{brown}{\phi} = hard$}{
    $\textcolor{violet}{D} \gets \text{GetMaxKeys}(C_{distractor}, \textcolor{orange}{N})$\;
}
\end{algorithm*}

%% file: Tables/models.tex
\begin{table*}[ht]
\centering
\resizebox{1\textwidth}{!}{%
\begin{tabular}{llllll}
\hlineB{2.7}
\textbf{Scale} & \textbf{Model} & \textbf{Parameters} & \textbf{Organisation} & \textbf{Release Date} & \textbf{Hugging Face} \\ \hline

\multicolumn{1}{l|}{\multirow{16}{*}{$\leq$3B}} & Llama-3.2-3B-Instruct & 3.2B & Meta & Sep 25, 2024 & \href{https://huggingface.co/meta-llama/Llama-3.2-3B-Instruct}{meta-llama/Llama-3.2-3B-Instruct} \\
\multicolumn{1}{l|}{} & Llama-3.2-1B-Instruct & 1.2B & Meta & Sep 25, 2024 &
\href{https://huggingface.co/meta-llama/Llama-3.2-1B-Instruct}{meta-llama/Llama-3.2-1B-Instruct}\\ 
\multicolumn{1}{l|}{} & Qwen2.5-Coder-3B-Instruct & 3.1B & Alibaba & Nov 6, 2024  &
\href{https://huggingface.co/Qwen/Qwen2.5-Coder-3B-Instruct}{Qwen/Qwen2.5-Coder-3B-Instruct}\\
\multicolumn{1}{l|}{} & Qwen2.5-Coder-1.5B-Instruct & 1.5B & Alibaba & Sep 18, 2024 &
\href{https://huggingface.co/Qwen/Qwen2.5-Coder-1.5B-Instruct}{Qwen/Qwen2.5-Coder-1.5B-Instruct}\\
\multicolumn{1}{l|}{} & Qwen2.5-3B-Instruct & 3.1B & Alibaba & Sep 18, 2024 &
\href{https://huggingface.co/Qwen/Qwen2.5-3B-Instruct}{Qwen/Qwen2.5-3B-Instruct}\\
\multicolumn{1}{l|}{} & Qwen2.5-1.5B-Instruct & 1.5B & Alibaba & Sep 18, 2024 &
\href{https://huggingface.co/Qwen/Qwen2.5-1.5B-Instruct}{Qwen/Qwen2.5-1.5B-Instruct}\\ 
\multicolumn{1}{l|}{} & deepseek-coder-1.3b-instruct & 1.3B & DeepSeek & Oct 30, 2023 &
\href{https://huggingface.co/deepseek-ai/deepseek-coder-1.3b-instruct}{deepseek-ai/deepseek-coder-1.3b-instruct}\\
\end{tabular}
}
\caption{List of benchmarked models.}
\label{table:model}
\end{table*}

%% file: Tables/ACR.tex
\begin{table*}[ht]
\centering
\resizebox{0.95\textwidth}{!}{%
\begin{tabular}{llcccccccccc}
\hlineB{2.7}
\textbf{Scale} &\textbf{Model} &\textbf{C} &\textbf{C++} &\textbf{CSharp} &\textbf{Go} &\textbf{Java} &\textbf{JavaScript} &\textbf{PHP} &\textbf{Python} &\textbf{Ruby}&\textbf{Overall}\\ \hline

\multicolumn{1}{l|}{\multirow{16}{*}{$\leq$3B}} &\cellcolor{gray!20}Llama-3.2-3B-Instruct &\cellcolor{gray!20}23.0 &\cellcolor{gray!20}18.0 &\cellcolor{gray!20}22.0 &\cellcolor{gray!20}21.0  &\cellcolor{gray!20}29.0 &\cellcolor{gray!20}24.0 &\cellcolor{gray!20}36.0 &\cellcolor{gray!20}\underline{33.0} &\cellcolor{gray!20}27.0 &\cellcolor{gray!20}25.9 \\
\multicolumn{1}{l|}{} & Llama-3.2-1B-Instruct  &4.0 &7.0 &1.0 &9.0 &11.0 &2.0 &9.0 &5.0 &3.0 &5.7 \\
\multicolumn{1}{l|}{} &\cellcolor{gray!20}Qwen2.5-Coder-3B-Instruct  &\cellcolor{gray!20}\underline{24.0} &\cellcolor{gray!20}\underline{28.0} &\cellcolor{gray!20}\underline{27.0} &\cellcolor{gray!20}\underline{30.0} &\cellcolor{gray!20}\underline{30.0} &\cellcolor{gray!20}25.0 &\cellcolor{gray!20}\underline{42.0} &\cellcolor{gray!20}30.0 &\cellcolor{gray!20}\underline{37.0} & \cellcolor{gray!20}\underline{30.3} \\
\multicolumn{1}{l|}{} & Qwen2.5-Coder-1.5B-Instruct  &11.0 &16.0  &19.0 &23.0 &23.0 &19.0 &36.0 &23.0 &21.0 &21.2 \\
\multicolumn{1}{l|}{} &\cellcolor{gray!20}Qwen2.5-3B-Instruct  &\cellcolor{gray!20}15.0 &\cellcolor{gray!20}16.0 &\cellcolor{gray!20}12.0 &\cellcolor{gray!20}24.0 &\cellcolor{gray!20}18.0 &\cellcolor{gray!20}\underline{26.0} &\cellcolor{gray!20}31.0 &\cellcolor{gray!20}24.0 &\cellcolor{gray!20}26.0 &\cellcolor{gray!20}21.3 \\
\multicolumn{1}{l|}{} & Qwen2.5-1.5B-Instruct  &14.0 &17.0 &12.0 &16.0 &24.0 &19.0 &27.0 &23.0 &28.0 &20.0 \\
\multicolumn{1}{l|}{} &\cellcolor{gray!20}deepseek-coder-1.3b-instruct  &\cellcolor{gray!20}7.0 &\cellcolor{gray!20}8.0 &\cellcolor{gray!20}10.0 &\cellcolor{gray!20}13.0 &\cellcolor{gray!20}13.0 &\cellcolor{gray!20}6.0 &\cellcolor{gray!20}16.0 &\cellcolor{gray!20}16.0 &\cellcolor{gray!20}16.0 &\cellcolor{gray!20}11.7 \\
\end{tabular}
}
\caption{Automated Code Refinement - Exact Match Rate (\%).}
\label{table:acr}
\end{table*}

%% file: Tables/CTR.tex
\begin{table*}[ht]
\centering
\resizebox{1\textwidth}{!}{%
\begin{tabular}{llccccccccccc}
\hlineB{2.7}
\textbf{Scale} &\textbf{Model} & \textbf{PPA (Random = 33.3)} &\textbf{C} &\textbf{C++} &\textbf{CSharp} &\textbf{Go} &\textbf{Java} &\textbf{JavaScript} &\textbf{PHP} &\textbf{Python} &\textbf{Ruby}&\textbf{Overall}\\ \hline

\multicolumn{1}{l|}{\multirow{16}{*}{$\leq$3B}} &\cellcolor{gray!20}Llama-3.2-3B-Instruct &\cellcolor{gray!20}99.2 (+65.8) &\cellcolor{gray!20}\underline{73.0} &\cellcolor{gray!20}\underline{80.0} &\cellcolor{gray!20}73.0 &\cellcolor{gray!20}\underline{80.0}  &\cellcolor{gray!20}82.0 &\cellcolor{gray!20}79.0 &\cellcolor{gray!20}\underline{85.0} &\cellcolor{gray!20}72.0 &\cellcolor{gray!20}85.0 &\cellcolor{gray!20}\underline{78.8} \\

\multicolumn{1}{l|}{} & Llama-3.2-1B-Instruct  &98.3 (+65.0) &72.0 &76.0 &72.0 &77.0 &79.0 &77.0 &80.0 &73.0 &79.0 &76.1 \\

\multicolumn{1}{l|}{} &\cellcolor{gray!20}Qwen2.5-Coder-3B-Instruct  &\cellcolor{gray!20}98.6 (+65.3) &\cellcolor{gray!20}72.0 &\cellcolor{gray!20}76.0 &\cellcolor{gray!20}72.0 &\cellcolor{gray!20}\underline{80.0}  &\cellcolor{gray!20}81.0 &\cellcolor{gray!20}\underline{80.0} &\cellcolor{gray!20}81.0 &\cellcolor{gray!20}71.0 &\cellcolor{gray!20}86.0 &\cellcolor{gray!20}77.7 \\

\multicolumn{1}{l|}{} & Qwen2.5-Coder-1.5B-Instruct  &95.7 (+62.3) &67.0 &74.0 &67.0 &75.0 &76.0 &73.0 &76.0 &65.0 &81.0 &72.7 \\

\multicolumn{1}{l|}{} &\cellcolor{gray!20}Qwen2.5-3B-Instruct  &\cellcolor{gray!20}94.0 (+60.6) &\cellcolor{gray!20}70.0 &\cellcolor{gray!20}76.0 &\cellcolor{gray!20}64.0 &\cellcolor{gray!20}72.0  &\cellcolor{gray!20}72.0 &\cellcolor{gray!20}64.0 &\cellcolor{gray!20}68.0 &\cellcolor{gray!20}59.0 &\cellcolor{gray!20}74.0 &\cellcolor{gray!20}68.8 \\

\multicolumn{1}{l|}{} & Qwen2.5-1.5B-Instruct  &97.7 (+64.4) &71.0 &79.0 &70.0 &\underline{80.0} &\underline{83.0} &78.0 &82.0 &\underline{74.0} &\underline{\textbf{87.0}} & 78.2 \\
\end{tabular}
}
\caption{Change Type Recognition - Proportion of Plurarity Agreement and Invariant Accuracy (\%).}
\label{table:ctr}
\end{table*}

%% file: Tables/CL_easy.tex
\begin{table*}[ht]
\centering
\resizebox{1\textwidth}{!}{%
\begin{tabular}{llccccccccccc}
\hlineB{2.7}
\textbf{Scale} &\textbf{Model} & \textbf{PPA (Random = 25.0)} &\textbf{C} &\textbf{C++} &\textbf{CSharp} &\textbf{Go} &\textbf{Java} &\textbf{JavaScript} &\textbf{PHP} &\textbf{Python} &\textbf{Ruby}&\textbf{Overall}\\ \hline

\multicolumn{1}{l|}{\multirow{16}{*}{$\leq$3B}} &\cellcolor{gray!20}Llama-3.2-3B-Instruct &\cellcolor{gray!20}47.3 (+22.3) &\cellcolor{gray!20}2.0 &\cellcolor{gray!20}0.0 &\cellcolor{gray!20}0.0 &\cellcolor{gray!20}0.0  &\cellcolor{gray!20}1.0 &\cellcolor{gray!20}0.0 &\cellcolor{gray!20}2.0 &\cellcolor{gray!20}1.0 &\cellcolor{gray!20}1.0 &\cellcolor{gray!20}0.8 \\

\multicolumn{1}{l|}{} & Llama-3.2-1B-Instruct  &32.9 (+7.9) &0.0 &0.0 &0.0 &0.0 &0.0 &0.0 &0.0 &0.0 &0.0 &0.0 \\

\multicolumn{1}{l|}{} &\cellcolor{gray!20}Qwen2.5-Coder-3B-Instruct  &\cellcolor{gray!20}49.6 (+24.6) &\cellcolor{gray!20}0.0 &\cellcolor{gray!20}4.0 &\cellcolor{gray!20}1.0 &\cellcolor{gray!20}2.0  &\cellcolor{gray!20}2.0 &\cellcolor{gray!20}0.0 &\cellcolor{gray!20}2.0 &\cellcolor{gray!20}2.0 &\cellcolor{gray!20}3.0 &\cellcolor{gray!20}1.8 \\

\multicolumn{1}{l|}{} & Qwen2.5-Coder-1.5B-Instruct  &25.7 (+0.7) &0.0 &0.0 &0.0 &0.0 &0.0 &0.0 &0.0 &0.0 &0.0 &0.0 \\

\multicolumn{1}{l|}{} &\cellcolor{gray!20}Qwen2.5-3B-Instruct &\cellcolor{gray!20}82.4 (+57.4) &\cellcolor{gray!20}\underline{41.0} &\cellcolor{gray!20}\underline{34.0} &\cellcolor{gray!20}\underline{45.0} &\cellcolor{gray!20}\underline{39.0}  &\cellcolor{gray!20}\underline{45.0} &\cellcolor{gray!20}\underline{31.0} &\cellcolor{gray!20}46.0 &\cellcolor{gray!20}37.0 &\cellcolor{gray!20}\underline{36.0} &\cellcolor{gray!20}\underline{39.3} \\

\multicolumn{1}{l|}{} & Qwen2.5-1.5B-Instruct &57.6 (+32.6) &2.0 &1.0 &2.0 &1.0 &3.0 &1.0 &4.0 &1.0 &1.0 &1.8 \\
\end{tabular}
}
\caption{Change Localisation (Easy) - Proportion of Plurarity Agreement and Invariant Accuracy (\%).}
\label{table:cl_easy}
\end{table*}

%% file: Tables/CL_hard.tex
\begin{table*}[ht]
\centering
\resizebox{1\textwidth}{!}{%
\begin{tabular}{llccccccccccc}
\hlineB{2.7}
\textbf{Scale} &\textbf{Model} & \textbf{PPA (Random = 25.0)} &\textbf{C} &\textbf{C++} &\textbf{CSharp} &\textbf{Go} &\textbf{Java} &\textbf{JavaScript} &\textbf{PHP} &\textbf{Python} &\textbf{Ruby}&\textbf{Overall}\\ \hline

\multicolumn{1}{l|}{\multirow{16}{*}{$\leq$3B}} &\cellcolor{gray!20}Llama-3.2-3B-Instruct &\cellcolor{gray!20}45.8 (+20.8) &\cellcolor{gray!20}0.0 &\cellcolor{gray!20}1.0 &\cellcolor{gray!20}0.0 &\cellcolor{gray!20}0.0 &\cellcolor{gray!20}0.0 &\cellcolor{gray!20}1.0 &\cellcolor{gray!20}1.0 &\cellcolor{gray!20}0.0 &\cellcolor{gray!20}1.0 &\cellcolor{gray!20}0.4 \\

\multicolumn{1}{l|}{} & Llama-3.2-1B-Instruct &31.3 (+6.3) &0.0 &0.0 &0.0 &0.0 &0.0 &0.0 &0.0 &0.0 &0.0 &0.0\\

\multicolumn{1}{l|}{} &\cellcolor{gray!20}Qwen2.5-Coder-3B-Instruct  &\cellcolor{gray!20}52.1 (+27.1) &\cellcolor{gray!20}0.0 &\cellcolor{gray!20}3.0 &\cellcolor{gray!20}0.0 &\cellcolor{gray!20}2.0 &\cellcolor{gray!20}2.0 &\cellcolor{gray!20}1.0 &\cellcolor{gray!20}3.0 &\cellcolor{gray!20}2.0 &\cellcolor{gray!20}3.0 &\cellcolor{gray!20}1.8\\

\multicolumn{1}{l|}{} & Qwen2.5-Coder-1.5B-Instruct  &27.0 (+2.0) & 0.0 & 0.0 & 0.0 & 0.0 & 0.0 & 0.0 & 0.0 & 0.0 & 0.0 & 0.0 \\

\multicolumn{1}{l|}{} &\cellcolor{gray!20}Qwen2.5-3B-Instruct  &\cellcolor{gray!20}78.0 (+53.0) &\cellcolor{gray!20}\underline{34.0} &\cellcolor{gray!20}\underline{27.0} &\cellcolor{gray!20}\underline{39.0} &\cellcolor{gray!20}31.0 &\cellcolor{gray!20}\underline{39.0} &\cellcolor{gray!20}\underline{28.0} &\cellcolor{gray!20}45.0 &\cellcolor{gray!20}34.0 &\cellcolor{gray!20}\underline{38.0} &\cellcolor{gray!20}\underline{35.0} \\

\multicolumn{1}{l|}{} & Qwen2.5-1.5B-Instruct  &54.5 (+29.5) & 0.0 & 1.0 & 1.0 & 1.0 & 2.0 & 1.0 & 4.0 & 0.0 & 1.0 & 1.2 \\
\end{tabular}
}
\caption{Change Localisation (Hard) - Proportion of Plurarity Agreement and Invariant Accuracy (\%).}
\label{table:cl_hard}
\end{table*}

%% file: Tables/SI_easy.tex
\begin{table*}[ht]
\centering
\resizebox{1\textwidth}{!}{%
\begin{tabular}{llccccccccccc}
\hlineB{2.7}
\textbf{Scale} &\textbf{Model} & \textbf{PPA (Random = 25.0)} &\textbf{C} &\textbf{C++} &\textbf{CSharp} &\textbf{Go} &\textbf{Java} &\textbf{JavaScript} &\textbf{PHP} &\textbf{Python} &\textbf{Ruby}&\textbf{Overall}\\ \hline

\multicolumn{1}{l|}{\multirow{16}{*}{$\leq$3B}} &\cellcolor{gray!20}Llama-3.2-3B-Instruct &\cellcolor{gray!20}49.3 (+24.3) &\cellcolor{gray!20}3.8 & \cellcolor{gray!20}5.9 & \cellcolor{gray!20}11.7 & \cellcolor{gray!20}16.3 & \cellcolor{gray!20}3.5 & \cellcolor{gray!20}7.0 & \cellcolor{gray!20}7.8 & \cellcolor{gray!20}9.0 & \cellcolor{gray!20}23.9 & \cellcolor{gray!20}9.9 \\

\multicolumn{1}{l|}{} & Llama-3.2-1B-Instruct &31.5 (+6.5) & 2.5 & 1.2 & 1.3 & 3.5 & 1.2 & 2.3 & 1.1 & 5.1 & 1.1 & 2.1 \\

\multicolumn{1}{l|}{} &\cellcolor{gray!20}Qwen2.5-Coder-3B-Instruct &\cellcolor{gray!20}44.5 (+19.5)  &\cellcolor{gray!20}11.3 & \cellcolor{gray!20}11.8 & \cellcolor{gray!20}15.6 & \cellcolor{gray!20}11.6 & \cellcolor{gray!20}17.4 & \cellcolor{gray!20}7.0 & \cellcolor{gray!20}16.7 & \cellcolor{gray!20}15.4 & \cellcolor{gray!20}3.4 & \cellcolor{gray!20}12.2\\

\multicolumn{1}{l|}{} & Qwen2.5-Coder-1.5B-Instruct &42.7 (+17.7) & 5.0 & 5.9 & 5.2 & 2.3 & 2.3 & 4.7 & 10.0 & 10.3 & 5.7 & 5.7 \\

\multicolumn{1}{l|}{} &\cellcolor{gray!20}Qwen2.5-3B-Instruct &\cellcolor{gray!20}82.5 (+57.5)  &\cellcolor{gray!20}50.0 & \cellcolor{gray!20}\underline{61.2} & \cellcolor{gray!20}51.9 & \cellcolor{gray!20}53.5 & \cellcolor{gray!20}59.3 & \cellcolor{gray!20}57.0 & \cellcolor{gray!20}\underline{62.2} & \cellcolor{gray!20}51.3 & \cellcolor{gray!20}51.1 & \cellcolor{gray!20}55.3\\

\multicolumn{1}{l|}{} & Qwen2.5-1.5B-Instruct &85.5 (+60.5)  & 32.5 & 52.9 & 46.8 & 40.7 & 44.2 & 44.2 & 57.8 & 39.7 & 52.3 & 45.7 \\
\end{tabular}
}
\caption{Solution Identification (Easy) - Proportion of Plurarity Agreement and Invariant Accuracy (\%).}
\label{table:si_easy}
\end{table*}

%% file: Tables/SI_hard.tex
\begin{table*}[ht]
\centering
\resizebox{1\textwidth}{!}{%
\begin{tabular}{llccccccccccc}
\hlineB{2.7}
\textbf{Scale} &\textbf{Model} & \textbf{PPA (Random = 25.0)} &\textbf{C} &\textbf{C++} &\textbf{CSharp} &\textbf{Go} &\textbf{Java} &\textbf{JavaScript} &\textbf{PHP} &\textbf{Python} &\textbf{Ruby}&\textbf{Overall}\\ \hline

\multicolumn{1}{l|}{\multirow{16}{*}{$\leq$3B}} &\cellcolor{gray!20}Llama-3.2-3B-Instruct &\cellcolor{gray!20}48.6 (+23.6) &\cellcolor{gray!20}5.0 & \cellcolor{gray!20}4.7 & \cellcolor{gray!20}7.8 & \cellcolor{gray!20}16.3 & \cellcolor{gray!20}1.2 & \cellcolor{gray!20}7.0 & \cellcolor{gray!20}6.7 & \cellcolor{gray!20}7.7 & \cellcolor{gray!20}12.5 & \cellcolor{gray!20}7.6 \\

\multicolumn{1}{l|}{} & Llama-3.2-1B-Instruct &31.2 (+6.2)  & 0.0 & 1.2 & 0.0 & 1.2 & 0.0 & 0.0 & 0.0 & 3.8 & 1.1 & 0.8\\

\multicolumn{1}{l|}{} &\cellcolor{gray!20}Qwen2.5-Coder-3B-Instruct &\cellcolor{gray!20}41.2 (+16.2)  &\cellcolor{gray!20}5.0 & \cellcolor{gray!20}4.7 & \cellcolor{gray!20}6.5 & \cellcolor{gray!20}11.6 & \cellcolor{gray!20}11.6 & \cellcolor{gray!20}11.6 & \cellcolor{gray!20}10.0 & \cellcolor{gray!20}9.0 & \cellcolor{gray!20}2.2 & \cellcolor{gray!20}8.0\\

\multicolumn{1}{l|}{} & Qwen2.5-Coder-1.5B-Instruct &41.1 (+16.1)  & 3.8 & 1.2 & 1.3 & 4.7 & 1.2 & 0.0 & 3.3 & 6.4 & 4.6 & 2.9\\

\multicolumn{1}{l|}{} &\cellcolor{gray!20}Qwen2.5-3B-Instruct &\cellcolor{gray!20}79.7 (+54.7)  &\cellcolor{gray!20}\underline{51.3} & \cellcolor{gray!20}\underline{50.6} & \cellcolor{gray!20}36.4 & \cellcolor{gray!20}\underline{47.7} & \cellcolor{gray!20}48.8 & \cellcolor{gray!20}45.4 & \cellcolor{gray!20}45.6 & \cellcolor{gray!20}41.0 & \cellcolor{gray!20}39.8 & \cellcolor{gray!20}45.2 \\

\multicolumn{1}{l|}{} & Qwen2.5-1.5B-Instruct &82.1 (+57.1)  &28.8 & 45.9 & 39.0 & 29.1 & 48.8 & 34.9 & 43.3 & 26.9 & 43.2 & 37.8\\
\end{tabular}
}
\caption{Solution Identification (Hard) - Proportion of Plurarity Agreement and Invariant Accuracy (\%).}
\label{table:si_hard}
\end{table*}

%% file: acl_latex.bbl
\begin{thebibliography}{37}
\providecommand{\natexlab}[1]{#1}

\bibitem[{Austin et~al.(2021)Austin, Odena, Nye, Bosma, Michalewski, Dohan, Jiang, Cai, Terry, Le, and Sutton}]{MBPP}
Jacob Austin, Augustus Odena, Maxwell~I. Nye, Maarten Bosma, Henryk Michalewski, David Dohan, Ellen Jiang, Carrie~J. Cai, Michael Terry, Quoc~V. Le, and Charles Sutton. 2021.
\newblock \href {https://arxiv.org/abs/2108.07732} {Program synthesis with large language models}.
\newblock \emph{CoRR}, abs/2108.07732.

\bibitem[{Baltes and Ralph(2022)}]{baltes2022sampling}
Sebastian Baltes and Paul Ralph. 2022.
\newblock \href {https://doi.org/10.1007/S10664-021-10072-8} {Sampling in software engineering research: a critical review and guidelines}.
\newblock \emph{Empirical Software Engineering}, 27(4):94.

\bibitem[{Chen et~al.(2021)Chen, Tworek, Jun, Yuan, de~Oliveira~Pinto, Kaplan, Edwards, Burda, Joseph, Brockman, Ray, Puri, Krueger, Petrov, Khlaaf, Sastry, Mishkin, Chan, Gray, Ryder, Pavlov, Power, Kaiser, Bavarian, Winter, Tillet, Such, Cummings, Plappert, Chantzis, Barnes, Herbert{-}Voss, Guss, Nichol, Paino, Tezak, Tang, Babuschkin, Balaji, Jain, Saunders, Hesse, Carr, Leike, Achiam, Misra, Morikawa, Radford, Knight, Brundage, Murati, Mayer, Welinder, McGrew, Amodei, McCandlish, Sutskever, and Zaremba}]{humaneval}
Mark Chen, Jerry Tworek, Heewoo Jun, Qiming Yuan, Henrique~Pond{\'{e}} de~Oliveira~Pinto, Jared Kaplan, Harri Edwards, Yuri Burda, Nicholas Joseph, Greg Brockman, Alex Ray, Raul Puri, Gretchen Krueger, Michael Petrov, Heidy Khlaaf, Girish Sastry, Pamela Mishkin, Brooke Chan, Scott Gray, Nick Ryder, Mikhail Pavlov, Alethea Power, Lukasz Kaiser, Mohammad Bavarian, Clemens Winter, Philippe Tillet, Felipe~Petroski Such, Dave Cummings, Matthias Plappert, Fotios Chantzis, Elizabeth Barnes, Ariel Herbert{-}Voss, William~Hebgen Guss, Alex Nichol, Alex Paino, Nikolas Tezak, Jie Tang, Igor Babuschkin, Suchir Balaji, Shantanu Jain, William Saunders, Christopher Hesse, Andrew~N. Carr, Jan Leike, Joshua Achiam, Vedant Misra, Evan Morikawa, Alec Radford, Matthew Knight, Miles Brundage, Mira Murati, Katie Mayer, Peter Welinder, Bob McGrew, Dario Amodei, Sam McCandlish, Ilya Sutskever, and Wojciech Zaremba. 2021.
\newblock \href {https://arxiv.org/abs/2107.03374} {Evaluating large language models trained on code}.
\newblock \emph{CoRR}, abs/2107.03374.

\bibitem[{Efstathiou and Spinellis(2018)}]{language_matters}
Vasiliki Efstathiou and Diomidis Spinellis. 2018.
\newblock \href {https://doi.org/10.1145/3183399.3183411} {Code review comments: language matters}.
\newblock In \emph{Proceedings of the 40th International Conference on Software Engineering: New Ideas and Emerging Results, {ICSE} {(NIER)} 2018, Gothenburg, Sweden, May 27 - June 03, 2018}, pages 69--72. {ACM}.

\bibitem[{Guo et~al.(2024)Guo, Cao, Xie, Liu, Li, Chen, and Peng}]{guo2024exploring}
Qi~Guo, Junming Cao, Xiaofei Xie, Shangqing Liu, Xiaohong Li, Bihuan Chen, and Xin Peng. 2024.
\newblock \href {https://doi.org/10.1145/3597503.3623306} {Exploring the potential of chatgpt in automated code refinement: An empirical study}.
\newblock In \emph{Proceedings of the 46th {IEEE/ACM} International Conference on Software Engineering, {ICSE} 2024, Lisbon, Portugal, April 14-20, 2024}, pages 34:1--34:13. {ACM}.

\bibitem[{Hendrycks et~al.(2021)Hendrycks, Burns, Basart, Zou, Mazeika, Song, and Steinhardt}]{MMLU}
Dan Hendrycks, Collin Burns, Steven Basart, Andy Zou, Mantas Mazeika, Dawn Song, and Jacob Steinhardt. 2021.
\newblock \href {https://openreview.net/forum?id=d7KBjmI3GmQ} {Measuring massive multitask language understanding}.
\newblock In \emph{9th International Conference on Learning Representations, {ICLR} 2021, Virtual Event, Austria, May 3-7, 2021}. OpenReview.net.

\bibitem[{Hu et~al.(2018)Hu, Li, Xia, Lo, and Jin}]{codecomment}
Xing Hu, Ge~Li, Xin Xia, David Lo, and Zhi Jin. 2018.
\newblock \href {https://doi.org/10.1145/3196321.3196334} {Deep code comment generation}.
\newblock In \emph{Proceedings of the 26th Conference on Program Comprehension, {ICPC} 2018, Gothenburg, Sweden, May 27-28, 2018}, pages 200--210. {ACM}.

\bibitem[{Jelinek et~al.(1977)Jelinek, Mercer, Bahl, and Baker}]{Jelinek1977PerplexityaMO}
Frederick Jelinek, Robert~L. Mercer, Lalit~R. Bahl, and Janet~M. Baker. 1977.
\newblock \href {https://api.semanticscholar.org/CorpusID:121680873} {Perplexity—a measure of the difficulty of speech recognition tasks}.
\newblock \emph{Journal of the Acoustical Society of America}, 62.

\bibitem[{Jiang et~al.(2017)Jiang, Armaly, and McMillan}]{commitgen}
Siyuan Jiang, Ameer Armaly, and Collin McMillan. 2017.
\newblock \href {https://doi.org/10.1109/ASE.2017.8115626} {Automatically generating commit messages from diffs using neural machine translation}.
\newblock In \emph{Proceedings of the 32nd {IEEE/ACM} International Conference on Automated Software Engineering, {ASE} 2017, Urbana, IL, USA, October 30 - November 03, 2017}, pages 135--146. {IEEE}.

\bibitem[{Jimenez et~al.(2024)Jimenez, Yang, Wettig, Yao, Pei, Press, and Narasimhan}]{swebench}
Carlos~E. Jimenez, John Yang, Alexander Wettig, Shunyu Yao, Kexin Pei, Ofir Press, and Karthik~R. Narasimhan. 2024.
\newblock \href {https://openreview.net/forum?id=VTF8yNQM66} {Swe-bench: Can language models resolve real-world github issues?}
\newblock In \emph{The Twelfth International Conference on Learning Representations, {ICLR} 2024, Vienna, Austria, May 7-11, 2024}. OpenReview.net.

\bibitem[{Kiyomaru et~al.(2024)Kiyomaru, Sugiura, Kawahara, and Kurohashi}]{kiyomaru_memorisation}
Hirokazu Kiyomaru, Issa Sugiura, Daisuke Kawahara, and Sadao Kurohashi. 2024.
\newblock \href {https://aclanthology.org/2024.inlg-main.45} {A comprehensive analysis of memorization in large language models}.
\newblock In \emph{Proceedings of the 17th International Natural Language Generation Conference, {INLG} 2024, Tokyo, Japan, September 23 - 27, 2024}, pages 584--596. ACL.

\bibitem[{Kwon et~al.(2023)Kwon, Li, Zhuang, Sheng, Zheng, Yu, Gonzalez, Zhang, and Stoica}]{vLLM}
Woosuk Kwon, Zhuohan Li, Siyuan Zhuang, Ying Sheng, Lianmin Zheng, Cody~Hao Yu, Joseph Gonzalez, Hao Zhang, and Ion Stoica. 2023.
\newblock \href {https://doi.org/10.1145/3600006.3613165} {Efficient memory management for large language model serving with pagedattention}.
\newblock In \emph{Proceedings of the 29th Symposium on Operating Systems Principles, {SOSP} 2023, Koblenz, Germany, October 23-26, 2023}, pages 611--626. {ACM}.

\bibitem[{Levenshtein(1966)}]{levenshtein1966binary}
VI~Levenshtein. 1966.
\newblock Binary codes capable of correcting deletions, insertions, and reversals.
\newblock \emph{Proceedings of the Soviet physics doklady}.

\bibitem[{Li et~al.(2022)Li, Lu, Guo, Duan, Jannu, Jenks, Majumder, Green, Svyatkovskiy, Fu, and Sundaresan}]{codereviewer}
Zhiyu Li, Shuai Lu, Daya Guo, Nan Duan, Shailesh Jannu, Grant Jenks, Deep Majumder, Jared Green, Alexey Svyatkovskiy, Shengyu Fu, and Neel Sundaresan. 2022.
\newblock \href {https://doi.org/10.1145/3540250.3549081} {Automating code review activities by large-scale pre-training}.
\newblock In \emph{Proceedings of the 30th {ACM} Joint European Software Engineering Conference and Symposium on the Foundations of Software Engineering, {ESEC/FSE} 2022, Singapore, Singapore, November 14-18, 2022}, pages 1035--1047. {ACM}.

\bibitem[{Lin et~al.(2024)Lin, Thongtanunam, Treude, and Charoenwet}]{Lin2024}
Hong~Yi Lin, Patanamon Thongtanunam, Christoph Treude, and Wachiraphan Charoenwet. 2024.
\newblock \href {https://doi.org/10.1145/3643991.3644910} {Improving automated code reviews: Learning from experience}.
\newblock In \emph{21st {IEEE/ACM} International Conference on Mining Software Repositories, {MSR} 2024, Lisbon, Portugal, April 15-16, 2024}, pages 278--283. {ACM}.

\bibitem[{Linares-V\'{a}squez et~al.(2014)Linares-V\'{a}squez, Klock, McMillan, Saban\'{e}, Poshyvanyk, and Gu\'{e}h\'{e}neuc}]{domainmatters}
Mario Linares-V\'{a}squez, Sam Klock, Collin McMillan, Aminata Saban\'{e}, Denys Poshyvanyk, and Yann-Ga\"{e}l Gu\'{e}h\'{e}neuc. 2014.
\newblock \href {https://doi.org/10.1145/2597008.2597144} {Domain matters: bringing further evidence of the relationships among anti-patterns, application domains, and quality-related metrics in java mobile apps}.
\newblock In \emph{22nd International Conference on Program Comprehension, {ICPC} 2014, Hyderabad, India, June 2-3, 2014}, pages 232--243. {ACM}.

\bibitem[{Liu et~al.(2025)Liu, Lin, and Thongtanunam}]{Liu2025Noisy}
Chunhua Liu, Hong~Yi Lin, and Patanamon Thongtanunam. 2025.
\newblock \href {https://doi.org/10.48550/arXiv.2502.02757} {Too noisy to learn: Enhancing data quality for code review comment generation}.
\newblock In \emph{22nd {IEEE/ACM} International Conference on Mining Software Repositories, {MSR} 2025, Ottawa, Canada, April 28-29, 2025}. {IEEE}.

\bibitem[{Liu et~al.(2019)Liu, Xia, Treude, Lo, and Li}]{prgen}
Zhongxin Liu, Xin Xia, Christoph Treude, David Lo, and Shanping Li. 2019.
\newblock \href {https://doi.org/10.1109/ASE.2019.00026} {Automatic generation of pull request descriptions}.
\newblock In \emph{34th {IEEE/ACM} International Conference on Automated Software Engineering, {ASE} 2019, San Diego, CA, USA, November 11-15, 2019}, pages 176--188. {IEEE}.

\bibitem[{Lu et~al.(2023)Lu, Yu, Li, Yang, and Zuo}]{llamareviewer}
Junyi Lu, Lei Yu, Xiaojia Li, Li~Yang, and Chun Zuo. 2023.
\newblock \href {https://doi.org/10.1109/ISSRE59848.2023.00026} {Llama-reviewer: Advancing code review automation with large language models through parameter-efficient fine-tuning}.
\newblock In \emph{34th {IEEE} International Symposium on Software Reliability Engineering, {ISSRE} 2023, Florence, Italy, October 9-12, 2023}, pages 647--658. {IEEE}.

\bibitem[{Papineni et~al.(2002)Papineni, Roukos, Ward, and Zhu}]{bleu}
Kishore Papineni, Salim Roukos, Todd Ward, and Wei{-}Jing Zhu. 2002.
\newblock \href {https://doi.org/10.3115/1073083.1073135} {Bleu: a method for automatic evaluation of machine translation}.
\newblock In \emph{Proceedings of the 40th Annual Meeting of the Association for Computational Linguistics, {ACL} 2002, Philadelphia, PA, {USA}, July 6-12, 2002}, pages 311--318. {ACL}.

\bibitem[{Paschali et~al.(2017)Paschali, Ampatzoglou, Bibi, Chatzigeorgiou, and Stamelos}]{reusability}
Maria~Eleni Paschali, Apostolos Ampatzoglou, Stamatia Bibi, Alexander Chatzigeorgiou, and Ioannis Stamelos. 2017.
\newblock \href {https://doi.org/10.1016/J.JSS.2017.09.009} {Reusability of open source software across domains: {A} case study}.
\newblock \emph{Journal of Systems and Software}, 134:211--227.

\bibitem[{Pornprasit and Tantithamthavorn(2024)}]{Pornprasit2024GPT3.5}
Chanathip Pornprasit and Chakkrit Tantithamthavorn. 2024.
\newblock \href {https://doi.org/10.1016/J.INFSOF.2024.107523} {Fine-tuning and prompt engineering for large language models-based code review automation}.
\newblock \emph{Information and Software Technology}, 175:107523.

\bibitem[{Rahman et~al.(2017)Rahman, Roy, and Kula}]{cr_usefulness}
Mohammad~Masudur Rahman, Chanchal~K. Roy, and Raula~Gaikovina Kula. 2017.
\newblock \href {https://doi.org/10.1109/MSR.2017.17} {Predicting usefulness of code review comments using textual features and developer experience}.
\newblock In \emph{Proceedings of the 14th International Conference on Mining Software Repositories, {MSR} 2017, Buenos Aires, Argentina, May 20-28, 2017}, pages 215--226. IEEE.

\bibitem[{Robinson and Wingate(2023)}]{MCSB}
Joshua Robinson and David Wingate. 2023.
\newblock \href {https://openreview.net/forum?id=yKbprarjc5B} {Leveraging large language models for multiple choice question answering}.
\newblock In \emph{The Eleventh International Conference on Learning Representations, {ICLR} 2023, Kigali, Rwanda, May 1-5, 2023}. OpenReview.net.

\bibitem[{Saha et~al.(2018)Saha, Lyu, Lam, Yoshida, and Prasad}]{bugsjar}
Ripon~K. Saha, Yingjun Lyu, Wing Lam, Hiroaki Yoshida, and Mukul~R. Prasad. 2018.
\newblock \href {https://doi.org/10.1145/3196398.3196473} {Bugs.jar: a large-scale, diverse dataset of real-world java bugs}.
\newblock In \emph{Proceedings of the 15th International Conference on Mining Software Repositories, {MSR} 2018, Gothenburg, Sweden, May 28-29, 2018}, pages 10--13. {ACM}.

\bibitem[{Sallou et~al.(2024)Sallou, Durieux, and Panichella}]{llmthreats}
June Sallou, Thomas Durieux, and Annibale Panichella. 2024.
\newblock \href {https://doi.org/10.1145/3639476.3639764} {Breaking the silence: the threats of using llms in software engineering}.
\newblock In \emph{Proceedings of the 2024 {ACM/IEEE} 44th International Conference on Software Engineering: New Ideas and Emerging Results, NIER@ICSE 2024, Lisbon, Portugal, April 14-20, 2024}, pages 102--106. {ACM}.

\bibitem[{Thongtanunam et~al.(2022)Thongtanunam, Pornprasit, and Tantithamthavorn}]{autotransform}
Patanamon Thongtanunam, Chanathip Pornprasit, and Chakkrit Tantithamthavorn. 2022.
\newblock \href {https://doi.org/10.1145/3510003.3510067} {Autotransform: Automated code transformation to support modern code review process}.
\newblock In \emph{44th {IEEE/ACM} 44th International Conference on Software Engineering, {ICSE} 2022, Pittsburgh, PA, USA, May 25-27, 2022}, pages 237--248. {ACM}.

\bibitem[{Tufano et~al.(2019)Tufano, Pantiuchina, Watson, Bavota, and Poshyvanyk}]{tufano2019}
Michele Tufano, Jevgenija Pantiuchina, Cody Watson, Gabriele Bavota, and Denys Poshyvanyk. 2019.
\newblock \href {https://doi.org/10.1109/ICSE.2019.00021} {On learning meaningful code changes via neural machine translation}.
\newblock In \emph{Proceedings of the 41st International Conference on Software Engineering, {ICSE} 2019, Montreal, QC, Canada, May 25-31, 2019}, pages 25--36. {IEEE} / {ACM}.

\bibitem[{Tufano et~al.(2024)Tufano, Dabic, Mastropaolo, Ciniselli, and Bavota}]{tufano2024code}
Rosalia Tufano, Ozren Dabic, Antonio Mastropaolo, Matteo Ciniselli, and Gabriele Bavota. 2024.
\newblock \href {https://doi.org/10.1109/TSE.2023.3348172} {Code review automation: Strengths and weaknesses of the state of the art}.
\newblock \emph{{IEEE} Transactions on Software Engineering}, 50(2):338--353.

\bibitem[{Tufano et~al.(2022)Tufano, Masiero, Mastropaolo, Pascarella, Poshyvanyk, and Bavota}]{tufano2022}
Rosalia Tufano, Simone Masiero, Antonio Mastropaolo, Luca Pascarella, Denys Poshyvanyk, and Gabriele Bavota. 2022.
\newblock \href {https://doi.org/10.1145/3510003.3510621} {Using pre-trained models to boost code review automation}.
\newblock In \emph{44th {IEEE/ACM} 44th International Conference on Software Engineering, {ICSE} 2022, Pittsburgh, PA, USA, May 25-27, 2022}, pages 2291--2302. {ACM}.

\bibitem[{Tufano et~al.(2021)Tufano, Pascarella, Tufano, Poshyvanyk, and Bavota}]{tufano2021}
Rosalia Tufano, Luca Pascarella, Michele Tufano, Denys Poshyvanyk, and Gabriele Bavota. 2021.
\newblock \href {https://doi.org/10.1109/ICSE43902.2021.00027} {Towards automating code review activities}.
\newblock In \emph{43rd {IEEE/ACM} International Conference on Software Engineering, {ICSE} 2021, Madrid, Spain, 22-30 May 2021}, pages 163--174. {IEEE}.

\bibitem[{Viggiato et~al.(2019)Viggiato, Oliveira, Figueiredo, Jamshidi, and K{\"{a}}stner}]{dompractices}
Markos Viggiato, Johnatan Oliveira, Eduardo Figueiredo, Pooyan Jamshidi, and Christian K{\"{a}}stner. 2019.
\newblock \href {https://doi.org/10.1109/ICGSE.2019.00013} {Understanding similarities and differences in software development practices across domains}.
\newblock In \emph{Proceedings of the 14th International Conference on Global Software Engineering, {ICGSE} 2019, Montreal, QC, Canada, May 25-31, 2019}, pages 74--84. {IEEE} / {ACM}.

\bibitem[{Wang et~al.(2025)Wang, Zhao, Qiang, Xi, Qin, and Liu}]{beyondanswers}
Haochun Wang, Sendong Zhao, Zewen Qiang, Nuwa Xi, Bing Qin, and Ting Liu. 2025.
\newblock \href {https://aclanthology.org/2025.coling-main.390/} {Llms may perform {MCQA} by selecting the least incorrect option}.
\newblock In \emph{Proceedings of the 31st International Conference on Computational Linguistics, {COLING} 2025, Abu Dhabi, UAE, January 19-24, 2025}, pages 5852--5862. {ACL}.

\bibitem[{Xu et~al.(2024)Xu, Wang, Fan, and Liu}]{xu_ngram}
Ruijie Xu, Zengzhi Wang, Run{-}Ze Fan, and Pengfei Liu. 2024.
\newblock \href {https://doi.org/10.48550/ARXIV.2404.18824} {Benchmarking benchmark leakage in large language models}.
\newblock \emph{CoRR}, abs/2404.18824.

\bibitem[{Yang et~al.(2023)Yang, Xu, Zhang, Zhang, and Bacchelli}]{evacrc}
Lanxin Yang, Jinwei Xu, Yifan Zhang, He~Zhang, and Alberto Bacchelli. 2023.
\newblock \href {https://doi.org/10.1145/3611643.3616245} {Evacrc: Evaluating code review comments}.
\newblock In \emph{Proceedings of the 31st {ACM} Joint European Software Engineering Conference and Symposium on the Foundations of Software Engineering, {ESEC/FSE} 2023, San Francisco, CA, USA, December 3-9, 2023}, pages 275--287. {ACM}.

\bibitem[{Zhu et~al.(2024)Zhu, Hao, He, Song, Yueyang, Zhang, Hu, Wei, Wang, and Lu}]{cleaneval}
Wenhong Zhu, Hongkun Hao, Zhiwei He, Yunze Song, Jiao Yueyang, Yumeng Zhang, Hanxu Hu, Yiran Wei, Rui Wang, and Hongyuan Lu. 2024.
\newblock \href {https://doi.org/10.18653/V1/2024.FINDINGS-NAACL.53} {{CLEAN-EVAL:} clean evaluation on contaminated large language models}.
\newblock In \emph{Findings of the Association for Computational Linguistics: {NAACL} 2024, Mexico City, Mexico, June 16-21, 2024}, pages 835--847. ACL.

\bibitem[{Zhuo et~al.(2025)Zhuo, Chien, Chim, Hu, Yu, Widyasari, Yusuf, Zhan, He, Paul, Brunner, GONG, Hoang, Zebaze, Hong, Li, Kaddour, Xu, Zhang, Yadav, Jain, Gu, Cheng, Liu, Liu, Wang, Lo, Hui, Muennighoff, Fried, Du, de~Vries, and Werra}]{bigcodebench}
Terry~Yue Zhuo, Vu~Minh Chien, Jenny Chim, Han Hu, Wenhao Yu, Ratnadira Widyasari, Imam Nur~Bani Yusuf, Haolan Zhan, Junda He, Indraneil Paul, Simon Brunner, Chen GONG, James Hoang, Armel~Randy Zebaze, Xiaoheng Hong, Wen-Ding Li, Jean Kaddour, Ming Xu, Zhihan Zhang, Prateek Yadav, Naman Jain, Alex Gu, Zhoujun Cheng, Jiawei Liu, Qian Liu, Zijian Wang, David Lo, Binyuan Hui, Niklas Muennighoff, Daniel Fried, Xiaoning Du, Harm de~Vries, and Leandro~Von Werra. 2025.
\newblock \href {https://openreview.net/forum?id=YrycTjllL0} {Bigcodebench: Benchmarking code generation with diverse function calls and complex instructions}.
\newblock In \emph{The Thirteenth International Conference on Learning Representations, {ICLR} 2025, Singapore, Singapore, Apr 24-28, 2025}. OpenReview.net.

\end{thebibliography}
